\documentclass[twocolumn]{IEEEtran}

\usepackage[mathscr]{eucal}
\usepackage[cmex10]{amsmath}
\usepackage{epsfig,epsf,psfrag}
\usepackage{amssymb,amsmath,amsthm,amsfonts,latexsym}
\usepackage{amsmath,graphicx,bm,xcolor,url}
\usepackage{fixltx2e}
\usepackage{array}
\usepackage{verbatim}
\usepackage{bm}
\usepackage{algorithmic, cite}
\usepackage{algorithm}
\usepackage{verbatim}
\usepackage{textcomp}
\usepackage{mathrsfs}
\usepackage{epstopdf}

\catcode`~=11 \def\UrlSpecials{\do\~{\kern -.15em\lower .7ex\hbox{~}\kern .04em}} \catcode`~=13 

\allowdisplaybreaks[4]
 
\newcommand{\nn}{\nonumber}

\newcommand{\calE}{\mathcal{E}}
\newcommand{\calF}{\mathcal{F}}
\newcommand{\calG}{\mathcal{G}}

\newcommand{\calK}{\mathcal{K}}
\newcommand{\calL}{\mathcal{L}}
\newcommand{\calM}{\mathcal{M}}

\newcommand{\calQ}{\mathcal{Q}}
\newcommand{\calR}{\mathcal{R}}
\newcommand{\calS}{\mathcal{S}}

\newcommand{\calV}{\mathcal{V}}

\newcommand{\calX}{\mathcal{X}}
\newcommand{\calY}{\mathcal{Y}}

\newcommand{\bx}{\mathbf{x}}

\newcommand{\rma}{\mathrm{a}}

\newcommand{\rmc}{\mathrm{c}}

\newcommand{\rmd}{\mathrm{d}}

\newcommand{\rme}{\mathrm{e}}
\newcommand{\rmE}{\mathrm{E}}

\newcommand{\rmm}{\mathrm{m}}

\newcommand{\rmQ}{\mathrm{Q}}
\newcommand{\rmr}{\mathrm{r}}

\newcommand{\rmT}{\mathrm{T}}

\newcommand{\rmx}{\mathrm{x}}

\newcommand{\bbE}{\mathbb{E}}

\newcommand{\bbN}{\mathbb{N}}

\newcommand{\bbP}{\mathbb{P}}

\newcommand{\bbR}{\mathbb{R}}

\newcommand{\bbZ}{\mathbb{Z}}

\DeclareMathAlphabet{\mathbsf}{OT1}{cmss}{bx}{n}
\DeclareMathAlphabet{\mathssf}{OT1}{cmss}{m}{sl}

\newcommand{\rvH}{\mathsf{H}}

\newcommand{\rvP}{\mathsf{P}}

\DeclareSymbolFont{bsfletters}{OT1}{cmss}{bx}{n}  
\DeclareSymbolFont{ssfletters}{OT1}{cmss}{m}{n}
\DeclareMathSymbol{\bsfGamma}{0}{bsfletters}{'000}
\DeclareMathSymbol{\ssfGamma}{0}{ssfletters}{'000}
\DeclareMathSymbol{\bsfDelta}{0}{bsfletters}{'001}
\DeclareMathSymbol{\ssfDelta}{0}{ssfletters}{'001}
\DeclareMathSymbol{\bsfTheta}{0}{bsfletters}{'002}
\DeclareMathSymbol{\ssfTheta}{0}{ssfletters}{'002}
\DeclareMathSymbol{\bsfLambda}{0}{bsfletters}{'003}
\DeclareMathSymbol{\ssfLambda}{0}{ssfletters}{'003}
\DeclareMathSymbol{\bsfXi}{0}{bsfletters}{'004}
\DeclareMathSymbol{\ssfXi}{0}{ssfletters}{'004}
\DeclareMathSymbol{\bsfPi}{0}{bsfletters}{'005}
\DeclareMathSymbol{\ssfPi}{0}{ssfletters}{'005}
\DeclareMathSymbol{\bsfSigma}{0}{bsfletters}{'006}
\DeclareMathSymbol{\ssfSigma}{0}{ssfletters}{'006}
\DeclareMathSymbol{\bsfUpsilon}{0}{bsfletters}{'007}
\DeclareMathSymbol{\ssfUpsilon}{0}{ssfletters}{'007}
\DeclareMathSymbol{\bsfPhi}{0}{bsfletters}{'010}
\DeclareMathSymbol{\ssfPhi}{0}{ssfletters}{'010}
\DeclareMathSymbol{\bsfPsi}{0}{bsfletters}{'011}
\DeclareMathSymbol{\ssfPsi}{0}{ssfletters}{'011}
\DeclareMathSymbol{\bsfOmega}{0}{bsfletters}{'012}
\DeclareMathSymbol{\ssfOmega}{0}{ssfletters}{'012}

\newcommand{\tilf}{\tilde{f}}

\newcommand{\tilg}{\tilde{g}}

\newcommand{\hatM}{\hat{M}}

\newcommand{\hatv}{\hat{v}}
\newcommand{\hatV}{\hat{V}}

\newcommand{\barQ}{\bar{Q}}

\def\fndot{\, \cdot \,}

\newcommand{\dotleq}{\stackrel{.}{\leq}}

\DeclareMathOperator*{\argmax}{arg\,max}

\newcommand{\bone}{\mathbf{1}}
\newcommand{\eps}{\varepsilon}

\newtheorem{theorem}{Theorem} 
\newtheorem{lemma}[theorem]{Lemma}

\newtheorem{proposition}[theorem]{Proposition}

\newtheorem{definition}{Definition}

\newcommand{\qednew}{\nobreak \ifvmode \relax \else
      \ifdim\lastskip<1.5em \hskip-\lastskip
      \hskip1.5em plus0em minus0.5em \fi \nobreak
      \vrule height0.75em width0.5em depth0.25em\fi}

\usepackage{xspace}
\usepackage[ colorlinks = true,
             linkcolor = blue,
             urlcolor  = blue,
             citecolor = red,
             anchorcolor = green,
]{hyperref}

\newcommand{\red}[1]{\textcolor{black}{#1}} 
\newcommand{\Red}[1]{\textcolor{black}{#1}} 
\newcommand{\blue}[1]{\textcolor{black}{#1}} 

\usepackage{cite}
\allowdisplaybreaks[1]
\newcommand{\hathh}{h}

\usepackage{amssymb}
\usepackage{pifont}
\newcommand{\cmark}{\ding{51}}%
\newcommand{\xmark}{\ding{55}}%
\usepackage{enumitem}

\begin{document}

\title{The Reliability Function of Variable-Length  Lossy Joint  Source-Channel Coding with Feedback} 
 
\author{Lan V.\ Truong, {\em Member, IEEE}  $\qquad$
        Vincent Y.~F.~Tan, {\em Senior Member, IEEE} \thanks{This paper was presented in part at the 2018 International Symposium on Information Theory (ISIT) in Vail, Colorado.}
\thanks{Lan V. Truong is with the Department of Computer Science, School of Computing, National University of Singapore (NUS). Vincent Y.~.F.~Tan is with the Department of Electrical and Computer Engineering and the Department of Mathematics, NUS. Emails: \url{truongvl@comp.nus.edu.sg};  \url{vtan@nus.edu.sg}} \thanks{ Copyright (c) 2019 IEEE. Personal use of this material is permitted.  However, permission to use this material for any other purposes must be obtained from the IEEE by sending a request to pubs-permissions@ieee.org.} 
}

\maketitle
 
\begin{abstract} 
We consider transmission of discrete memoryless sources (DMSes) across discrete memoryless channels (DMCs) using variable-length lossy source-channel codes with feedback. 
The reliability function (optimum error exponent) is shown to be equal to $\max\{0, B(1-R(D)/C)\},$ where $R(D)$ is the rate-distortion function of the source, $B$ is the maximum relative entropy between output distributions of the DMC, and $C$ is the Shannon capacity of the channel.  {We show that in this asymptotic regime, separate source-channel coding is, in fact, optimal.}
\end{abstract}   
\begin{IEEEkeywords}
Variable-length codes, Joint source-channel coding, Feedback, Reliability function. 
\end{IEEEkeywords}
 \section{Introduction}
The communication model  for discrete memoryless channel (DMCs) with feedback in which the  blocklength $\tau\in\bbN$ is a random variable whose expectation is over bounded by some positive real number $N\in\bbR_+$ was first proposed by Burnashev in a seminal work~\cite{Burnashev1976}. He demonstrated that the {\em reliability function} or {\em optimal error exponent}  for DMCs   with feedback improves dramatically over the no feedback case and the case where the blocklength is deterministic. This  class of  codes is known as {\em variable-length  codes with feedback}. The reliability function of a DMC with variable-length feedback admits a particularly simple expression 
$E_{\mathrm{Burn}}(R)= B (1-R/C )$ for all rates $0\le R\le C$,  
where $C$ is the capacity of the DMC and $B$ (usually written as $C_1$ in the literature) is  the relative entropy between conditional output distributions of the two most ``most distinguishable'' channel input symbols~\cite{Burnashev1976}. In this paper, we consider variable-length transmission of a discrete memoryless source (DMS) over a DMC with feedback under an excess-distortion constraint. Different from the recent elegant work of Kostina, Polyanskiy, and  Verd\'u~\cite{kost17} which considers the minimum expected delay (length) of such variable-length joint source-channel codes with feedback under a non-vanishing excess-distortion probability, we are interested in finding the optimal excess-distortion exponent  (reliability function) of such codes.
\subsection{Related Works} \label{sec:related}

Source-channel codes with deterministic (non-random) source and channel  block lengths were first introduced by Shannon~\cite{Shannon59b} in 1959. By Shannon's fundamental source and channel coding theorems, transmission with vanishing probability of error (or probability of excess distortion) is possible whenever the source entropy (or rate-distortion function in the lossy case) is less than the channel capacity. Gallager~\cite{gallagerIT} and Jelinek~\cite{jelinek} indicated that \red{joint} (i.e., not separate) source-channel coding can lead to a larger error exponent, which means that the separation rule for joint source-channel coding does not hold from the error exponent perspective. For $R\geq R_{\rmc\rmr}$ (where $R_{\mathrm{cr}}$ is the critical rate of the DMC),  Csisz\'ar~\cite{Csi80} later proved that the optimal error exponent for lossless joint source-channel coding is equal to $\min_{R}\{e(R)+E(R)\}$, where $e(R)$ and $E(R)$ are reliability functions of the DMS and DMC, respectively. The achievable joint source-channel coding scheme which was proposed in~\cite{Csi80} is a {\em universal}   code, i.e., the coding scheme does not depend on knowledge of the DMS or the DMC.

Wang, Ingber, and Kochman~\cite{wang12} recently proved that the no-excess-distortion probability has an exponential behavior for any lossy joint source-channel codes with fixed-length joint source-channel coding under the condition that $R(D)>C$ (assuming that the source and channel blocklengths are the same). Furthermore, they showed that the best exponential behavior (or strong converse exponent) is attainable by a separation-based scheme. The fact that separation is optimal for the the non-excess-distortion exponent    can be explained by observing the fact that the probability of non-excess-distortion can be approximated by product of the probability of non-excess-distortion in source coding and the probability of correct decoding in the channel coding phase under the condition that $R(D)>C$. Their achievable separation-based joint source-channel code is a universal joint source-channel code which is based on~\cite{Dueck1979}. 

By assuming side information available at decoder, the celebrated results of Shannon~\cite{Shannon48} and Slepian and Wolf~\cite{sw73} imply that almost lossless communication is possible using separate source and channel codes if $H(U|V)<C$ for a source $U$ with side information $V$ and a channel with capacity $C$. On the other hand, Shamai and Verd\'{u}~\cite{Shamai1995} proved that codes with $H(U|V )>C$ cannot exist even if joint source-channel coding is employed. Hence, for the problem of transmitting a DMS over a DMC   with side information available at decoder, separate source and channel coding is asymptotically optimal when a vanishingly small probability of error is allowed. Using properties of the Lov\'asz theta function, Nayak, Tuncel, and Rose~\cite{Nayak06} later showed that separate source and channel coding is asymptotically suboptimal in general for the problem of designing codes for zero-error transmission of a source through a channel when the receiver has side information about the source. They also derived conditions on sources and channels for the optimality of separate source-channel coding. 

Recently, Kostina, Polyanskiy, and Verd\'u~\cite{kost17} quantified the minimal average delay (code length) attainable by lossy   source-channel codes with feedback and concluded that such codes lead  to a significant improvement in the fundamental delay-distortion tradeoff. They showed that   separate source-channel coding fails to achieve these minimal average delays if a non-vanishing distortion probability is allowed. In addition, the authors also investigated the minimum energy required to reproduce $N$ source samples with a given fidelity after it is transmitted over a memoryless Gaussian channel, and they showed that the required minimum energy is reduced with feedback and an average power constraint.
\subsection{Main Contributions}

We show that for variable-length joint source-channel codes with feedback, the  optimal excess-distortion probability is (to first-order in the exponent) equal to  $\exp(- B(1-R(D)/C)N)$, where $N$ is the expectation of the  blocklength, $R(D)$ is the rate-distortion function of the DMS, $C$ is the capacity of the DMC, and $B$ (usually written as $C_1$ in the literature) is  the relative entropy between conditional output distributions of the two most ``most distinguishable'' channel input symbols~\cite{Burnashev1976}. Our technical contributions are twofold. 
\begin{enumerate}[leftmargin=*]
\item Our first contribution, the direct part, is to judiciously modify  Yamamoto-Itoh's coding scheme~\cite{YamamotoItoh1979} by combining it with known error exponent results in lossy source coding~\cite{Marton74} so that it becomes amenable to  joint source-channel coding for the DMC with feedback. We ensure that the so constructed code achieves the excess-distortion   exponent $B(1-R(D)/C)$. 
\item Our second and main contribution, the converse part, consists in providing several new and novel analytical arguments (e.g., Lemma~\ref{Berlinlem}) to upper bound the excess-distortion error exponents of variable-length   source-channel codes with feedback. Our proof techniques are based partly on Berlin {\em et al.}'s~\cite{Berlin2009a} simplified converse proof of Burnashev's  exponent~\cite{Burnashev1976}. The most interesting contribution for this part is the introduction and analysis of a new (and optimal) decoding rule that is amenable to lossy  joint source-channel coding problems. This rule is called the \emph{distortion-MAP rule}. The well-known MAP decoding rule can be considered as a special case of the the distortion-MAP rule when the  permitted distortion is equal to zero.
\end{enumerate}
\subsection{Organization of the Paper}
The rest of this paper is structured as follows. In Section~\ref{sec:dmc_bc}, we provide a precise problem statement for variable-length source-channel coding with feedback. We state the main result in Section \ref{sec:mainresult}. The achievability proof is provided in Section~\ref{sec:ach}, and the converse proof is provided in Section~\ref{sec:converse}. Technical derivations are relegated to the appendices.
 \subsection{Notational Conventions} 
We use $\ln x$ to denote the natural logarithm so information units throughout are in nats. The binary entropy function is defined as $h(x):=-x\ln x-(1-x)\ln(1-x)$ for $x\in [0,1]$. The minimum of two numbers $a$ and $b$ is denoted interchangeably as $\min\{a,b\}$ and $a\wedge b$. As is usual in information theory, $Z_i^j$ denotes the vector $(Z_i,Z_{i+1},\ldots, Z_j)$. In this paper, we also define $\alpha/0=\infty$ for all $\alpha \geq 0$ and $0 \times \infty=0$.
\section{Problem Setting}\label{sec:dmc_bc}

\subsection{Basic Definitions}

\begin{figure*}
\centering
\setlength{\unitlength}{.4mm}
\begin{picture}(350,75)
\put(50,00){\line(1,0){50}}
\put(50,30){\line(1,0){50}}
\put(50,00){\line(0,1){30}}
\put(100,00){\line(0,1){30}}

\put(00,15){\vector(1,0){50}}

\put(150,00){\line(1,0){50}}
\put(150,30){\line(1,0){50}}
\put(150,00){\line(0,1){30}}
\put(200,00){\line(0,1){30}}

\put(100,15){\vector(1,0){50}}

\put(250,00){\line(1,0){50}}
\put(250,30){\line(1,0){50}}
\put(250,00){\line(0,1){30}}
\put(300,00){\line(0,1){30}}

\put(200,15){\vector(1,0){50}}

\put(300,15){\vector(1,0){50}}

\put(5,20){$\{V_n\}_{n=1}^{\infty}$}
\put(70,13){$f_n$}

\put(50,60){$Y^{n-1}$}

\put(107,20){$\{X_n\}_{n=1}^{\infty}$}
\put(163,13){$P_{Y|X}$}

\put(205,20){$\{Y_n\}_{n=1}^{\infty}$}

\put(270,13){$g_n$}

\put(310,20){$ \hatV^N(\tau_N)$}

\put(240,15){\line(0,1){60}}
\put(240,75){\vector(-1,0){40}}

\put(163,73){$\mbox{Delay}$}
\put(150,75){\line(-1,0){75}}
\put(75,75){\vector(0,-1){45}}
\put(150,60){\line(1,0){50}}
\put(150,90){\line(1,0){50}}
\put(150,60){\line(0,1){30}}
\put(200,60){\line(0,1){30}}

\end{picture}
	\caption{Joint source-channel coding using variable-length codes with feedback. The stopping time is $\tau_N$.}
	\label{fig:ISCCJournal}
\end{figure*}
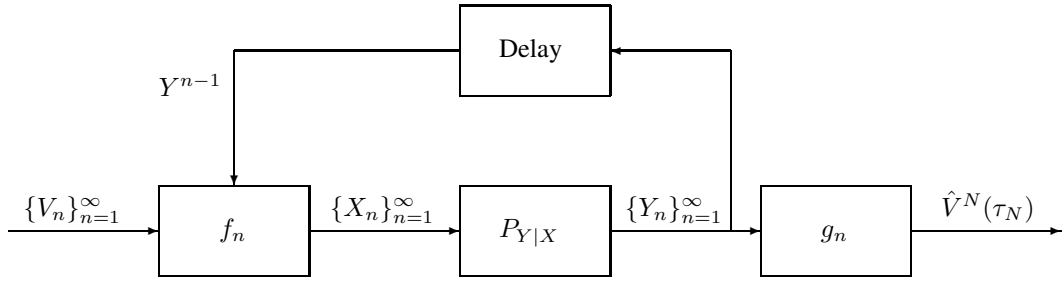

 Throughout, we let $\{V_n\}_{n=1}^{\infty}$ be DMS with distribution $P_V$ and taking values in a finite set $\calV$.
\begin{definition} 
\label{def1} A $(|\calV|^N,N)$-{\em variable-length joint source-channel code with feedback} for a DMC $P_{Y|X}$ (see Fig.~\ref{fig:ISCCJournal}), where $N$ is a positive integer, is defined by
\begin{itemize}[leftmargin=*]
\item A sequence of encoders $f_n: \calV^N \times \calY^{n-1} \to \calX, n\geq 1$, defining channel inputs
\begin{align}
X_n=f_n(V^N,Y^{n-1});\label{eqn:enc}
\end{align} 
\item A sequence of decoders $g_n: \calY^n \to \calV^N, n\geq 1$, each providing an estimate \red{$\hat{V}^N(n):=g_n(Y^n) \in \calV^N$} at time $n$;
\item An integer-valued random variable $\tau_N$ which is a stopping time of the filtration $\{\sigma(Y^n)\}_{n=0}^{\infty}$. 
\end{itemize}
\end{definition}
The final decision at the decoder is computed at  the stopping time $\tau_N$ as follows:
\begin{align}
\hat{V}^N({\tau_N}):=g_{\tau_N}(Y^{\tau_N}).
\end{align} 
The {\em excess-distortion probability} of the  coding scheme specified above is defined as
\begin{align}
\label{errordef}
\rvP_{\rmd}(N,D):=\bbP(d(\hat{V}^N({\tau_N}),V^N)>D),
\end{align} 
for some distortion measure $d: \calV^N \times \calV^N \to [0,+\infty)$ satisfying the following properties:   
\begin{align}
\label{measurepro1}
d(v^N,\hat{v}^N)&=\frac{1}{N}\sum_{i=1}^N d(v_i,\hat{v}_i),\\
\label{maxdist}
d_{\rmm\rma \rmx}&=\max_{(v,\hatv)\in \calV\times\calV} d(v,\hat{v}) <\infty,
\end{align} for any pair of  sequences $v^N \in \calV^N$ and $\hat{v}^N \in \calV^N$.
\begin{definition} \label{def:rel}
$E \in \bbR_+$ is an {\em achievable excess-distortion exponent  at distortion level $D$} if there exists a sequence of $(|\calV|^N,N)$-variable-length joint source-channel codes with feedback indexed by $N \in \bbN$  satisfying
\begin{align}
\label{eq8def}
\limsup_{N\to \infty} \frac{\bbE(\tau_N)}{N}&\leq 1,\\
\label{key1001}
\liminf_{N\to \infty}\,\,-\frac{\ln \rvP_{\rmd}(N,D)}{N} &\geq E.
\end{align}  The {\em excess-distortion reliability function}  using variable-length joint source-channel codes with feedback is 
\begin{align}
E^*(D):=\sup\{E:& \mbox{ is an achievable excess-distortion} \nn\\*
& \quad \mbox{exponent at distortion level $D$}\}.
\end{align} 
\end{definition}

%
%
%
%
%
%
%
%

\begin{definition} \label{def:info_q} Given a  DMC  $P_{Y|X}$,  define the channel parameters 
\begin{align}
\label{defineB}
B&:=\max_{x,x'\in \calX} D(P_{Y|X}(\cdot|x)\|P_{Y|X}(\cdot|x'))\\
\lambda &:=\min_{( x,y) \in \calX\times \calY}P_{Y|X}(y|x)\\
C&:=\max_{P_X} I(X;Y),
\end{align} 
Note that if $B<\infty$,  then $\lambda\in (0,1/2]$~\cite[Prop.~2]{Berlin2009a}. 
\end{definition}
In addition, define the {\em $d$-distortion ball} and the {\em rate-distortion function} respectively as
\begin{align}
\calS_D(v^N)&:=\{w^N \in \calV^N: d(w^N,v^N) \leq D\},\\
\label{distfunc}
R(Q,D)&:=\min_{P_{\hatV |V} :    \bbE_{Q\times P_{\hatV|V}}  [d(V,\hatV)] \leq D}  I(V;\hatV).
\end{align}
 If $Q=P_V$, we write $R(P_V,D)=R(D)$ for brevity.
\section{Main Result}\label{sec:mainresult}
\begin{theorem} \label{thm200} Assuming $B<\infty$, the following holds:
\begin{align}
\label{mainthm}
E^*(D)=\max\left\{0, B\left(1-\frac{R(D)}{C}\right)\right\}.
\end{align}
\end{theorem}
\begin{IEEEproof}
The proof is a combination of Propositions~\ref{achieve:proof} and~\ref{converse} in Sections~\ref{sec:ach} and~\ref{sec:converse}, respectively.
\end{IEEEproof}
Some remarks in order.
\begin{enumerate}[leftmargin=*]
\item If $D=0$, the problem reduces to (almost) lossless source coding and $R(D)|_{D=0}=H(P_V)$. If the source $P_V$ is uniformly distributed over $\calV$ and  with $R:=\log|\calV|$, then~\eqref{mainthm} reduces to Burnashev's exponent $E_{\mathrm{Burn}}(R)= \red{\max\{0,B (1-R/C )\}}$~\cite{Burnashev1976}, where  $R$ represents the rate of the channel code. 
\item   Since the   encoder  can use the same algorithm as the decoder to  detect  erroneous  source sequences (which are sequences in the message mode that have distance from the transmitted sequence  greater than $D$) by using feedback link, it can let the decoder know whether there is an error (by sending an ACK or NACK symbol). In addition,  by using the same decoding strategy, the encoder can also learn the decoded message at the decoder.
Therefore, the design of the  variable-length  joint source-channel code with feedback  is equivalent to the problem of designing  and subsequently concatenating  an  error exponent-optimal channel code and an  excess-distortion exponent-optimal lossy source code to minimize the retransmission probability. This intuitively means that the separation rule may be optimal for   variable-length source-channel coding with feedback  in the error exponents regime. Indeed, from the proof of Theorem~\ref{thm200}, we prove that the  separation is optimal for $R(D)<C$ in the asymptotic  regime of interest. In contrast,  Kostina, Polyanskiy, and Verd\'{u}~\cite{kost17} considered the non-vanishing error formalism for the same problem  and concluded that separation is not optimal. Without feedback, separation is also not optimal in the non-vanishing error regime~\cite{kost13}.  
\item For fixed-length   source-channel coding without feedback, Gallager~\cite{gallagerIT} and Jelinek~\cite{jelinek} indicated that joint source-channel coding leads to a larger error exponent than the separation scheme. More specifically, for $R\geq R_{\rmc\rmr}$,    Csisz\'ar~\cite{Csi80} proved that the optimal error exponent for lossless joint source-channel coding is equal to $\min_{R}\{e(R)+E(R)\}$, where $e(R)$ and $E(R)$ are  source and channel reliability functions (of the DMS and DMC)  respectively. This fact can be explained as follows. In Csisz\'ar's setup, the problem consists in designing a joint   source-channel code which has as large an error exponent as possible. The end-to-end error probability can be expressed as 
$\rvP_{\rme} = \rvP_{\rme,\mathrm{src}} + \rvP_{\rme,\mathrm{ch}}- \rvP_{\rme,\mathrm{src}}\times \rvP_{\rme,\mathrm{ch}}$; this indicates that a separate scheme is likely to be suboptimal as the smallest exponent  of the probabilities in this expression dominates. 
Indeed, for channels without feedback, there is no mechanism to detect errors. 
\item In contrast, in the strong converse exponent regime (still without feedback), Wang, Ingber, and Kochman~\cite{wang12} showed that separation is optimal. This is because the end-to-end correct decoding probability is 
$
\rvP_{\rmc} = \rvP_{\rmc,\mathrm{src}}  \times \rvP_{\rmc,\mathrm{ch}}, 
$ so the resultant exponent is the sum of the exponents of each probability. This hints at the fact that the source and channel coding can be designed independently and the overall system still performs optimally. \red{In our variable-length feedback setting, however, there is no strong converse~\cite{PPV11b}. Hence, separation trivially holds.} These discussions concerning optimality of separation in various settings are summarized in Table~\ref{tab:sep}.

\end{enumerate}

\begin{table*} 
\centering{\small
  \begin{tabular}{| c | c | c  |}
    \hline
    & No Feedback &   Variable-Length Feedback \\ \hline 
   Error Exponent &  \xmark~\cite{gallagerIT, jelinek, Csi80}   & \cmark $\,\, $Present work \\ \hline
   Strong Converse Exponent  & \cmark \cite{wang12} &  \cmark $\,\, $Implied by \cite{PPV11b} \\ \hline
   Fixed Error (Second-order) & \xmark~\cite{kost13} & \xmark~\cite{kost17}\\
    \hline
  \end{tabular}}\vspace{.05in}
  \caption{Table of whether separation is optimal for source-channel transmission in the error (and strong converse) exponents regime under various settings}
  \label{tab:sep}
\end{table*}
\section{Achievability Proof}\label{sec:ach}
\begin{definition}\cite[Chapter~2]{Csi97} Given a DMS which produces an i.i.d.\ sequence $V_1,V_2,\cdots,V_N   \sim P_V$, a $(|\calV|^N,N)$-block code $(\tilde{f}_{N},\tilde{g}_{N})$ for a source consists of
\begin{itemize}[leftmargin=*]
\item An encoding function $\tilde{f}_{N}: \calV^N \to \calR(\tilde{f}_{N})$,
 where $\calR(\tilde{f}_{N})$, the range of the encoding function $\tilde{f}_{N}$, is some finite set; 
\item A decoding function $\tilde{g}_{N}: \calR(\tilde{f}_{N}) \to \calV^N$.
\end{itemize}
\end{definition}
Define Marton's exponent~\cite{Marton74}
\begin{align} 
F(P_V,R,D):=\inf_{Q: R(Q,D)>R} D(Q\,\|\,P_V),
\end{align} 
\red{where $R(Q,D)$ is defined in~\eqref{distfunc}. We now state Marton's error exponent result for lossy source coding \cite{Marton74} \cite[Chapter 9]{Csi97}.}
\begin{lemma}\label{keycor}For any $\eps>0$ and $D\geq 0$, there exists a sequence of $(|\calV|^N,N)$-block codes $\{(\tilde{f}_{N},\tilde{g}_{N})\}_{N\geq 1}$ for the source $V^N\sim P_V^N$ such that
\begin{align}
|\calR(\tilde{f}_{N})| \leq \exp\left(N(R(D)+2\eps) \right)
\end{align} holds \Red{where $R(D)=R(P_V,D)$} and the probability of excess-distortion satisfies
\begin{align}
\label{zeroten}
\bbP(V^N \in \calL_N) \leq \exp\Big(-\frac{N}{2}F(P_V,R(D)+\eps,D)\Big) 
\end{align} for $N$ sufficiently large, where
\begin{align}
\calL_N:=\big\{v^N \in \calV^N: d(v^N,\tilde{g}_{N}(\tilde{f}_{N}(v^N)))>D \big\}. \label{eqn:defLN}
\end{align} 
\end{lemma}
\begin{proposition}[Achievability] \label{achieve:proof} The following inequality holds:
\begin{align}
\label{eqachieve}
E^*(D)\geq \max\left\{0, B\left(1-\frac{R(D)}{C}\right)\right\}.
\end{align}
\end{proposition}

\begin{figure}[t]
\centering
\setlength{\unitlength}{.4mm}
\begin{picture}(200, 70)
\put(0, 20){\vector(1, 0){80}}
\put(80, 20){\vector(-1, 0){80}}
\put(80, 20){\vector(1, 0){120}}
\put(200, 20){\vector(-1, 0){120}}
\put(0, 30){\line(1, 0){200}}
\put(0, 20){\line(0, 1){20}}
\put(80, 20){\line(0, 1){20}}
\put(200, 20){\line(0, 1){20}}
\put(0, 7){Length-$\gamma N$ msg mode}
\put(90, 7){Length-$(1-\gamma) N$ control mode}
\put(7, 47){One msg in $\calM$}
\put(13, 37){transmitted}
\put(100, 47){Control signals $\rmc$ and $\rme$ }
\put(120, 37){transmitted}
\put(80, 65){\vector(0, -1){20}}
\put(200, 65){\vector(0, -1){20}}
\put(70, 70){Feedback}
\put(170, 70){Feedback}
  \end{picture}
	\caption{A single  length-$N$ block in the Yamamoto-Itoh coding scheme which is repeated multiple times as described in Step~3 of the proof of Proposition~\ref{achieve:proof}.}
	\label{fig:YI1979}
\end{figure}
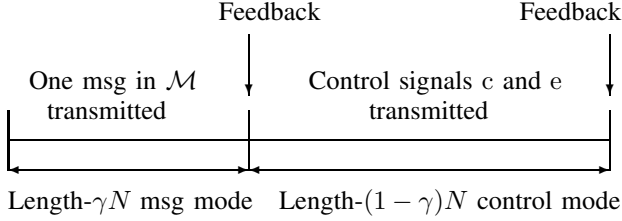

We now provide an overview of the proof strategy of the achievability part. In this part, we show that~\eqref{eqachieve} can be achieved by modifying the Yamamoto-Itoh coding scheme~\cite{YamamotoItoh1979}. The Yamamoto-Itoh coding scheme is a sequence of Yamamoto-Itoh coding blocks (YICBs) of length $N$. Each YICB  has two phases: a message phase and a control phase (see Fig.~\ref{fig:YI1979}). In the message phase, the encoder encodes the transmitted message by using a random codebook. It then  sends one of $|\calM|\in\bbN$ messages to the decoder {via the forward link of the channel}. The decoder decodes the transmitted message by using a maximum likelihood decoder. 
The decoder then sends the received signal $Y^n$ 
 through the noiseless feedback \red{link of the channel}. The encoder emulates the decoder to retrieve the decoded message at the decoder. In the control phase, the encoder sends the control signal $\rmc$ (ACK) if the decoder is correct at the message phase or the control signal $\rme$ (NACK) otherwise. The codewords for $\rmc$ and $\rme$ form a  repetition codebook. The decoder   decodes the control signal by using the decoder for the control phase of the YICB as in~\cite{YamamotoItoh1979}. 
 If the decoded result at the control phase is $\rmc$, the decoder declares the decoded message at the message phase to be correct and signals to the encoder to stop transmission via the feedback link of the channel. If the output of the decoder is $\rme$, the decoder discards the decoded message in the message phase. \Red{When the encoder knows that   $\rme$ is decoded at the end of a YICB, it resends the same message in the message phase of the next coding block.\footnote{Thanks to this retransmission mechanism, Nakibo\u{g}lu and Gallager also showed that errors occur independently after each repetition of the length-$N$ YICB as~\cite[Sec.~II.B]{Nak08} given a fixed transmitted message. Thus, the error probability is easy to bound by invoking properties of the geometric distribution.} This process continues until the encoder knows that $\rmc$  is decoded in the control phase of a particular YICB by its emulation of the decoder to decode the control message.} More precisely, our modification of the standard Yamamoto-Itoh coding scheme to become a variable-length joint source-channel code with feedback is detailed in Step 1 and analyzed in Steps 2 to 5 below. 

\begin{IEEEproof}[Proof of~\red{Proposition~\ref{achieve:proof}}] 
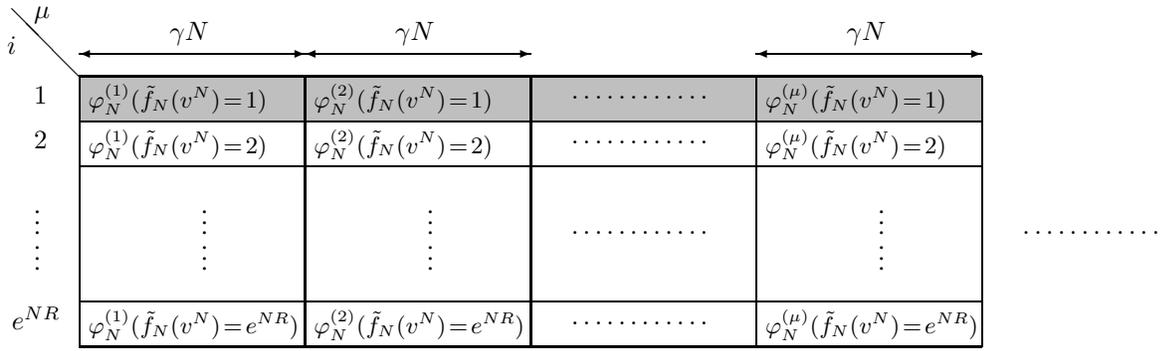
\begin{figure*}[t]
\centering
\setlength{\unitlength}{.6mm}
\begin{picture}(260,75)
\put(10,00){\line(1,0){200}}
\put(10,10){\line(1,0){100}}

\put(187,25){$\vdots$}
\put(87,25){$\vdots$}
\put(37,25){$\vdots$}
\put(37,17){$\vdots$}
\put(87,17){$\vdots$}
\put(187,17){$\vdots$}

\put(0,25){$\vdots$}
\put(0,17){$\vdots$}

\put(10,60){\line(1,0){200}}
\put(10,50){\line(1,0){100}}
\put(10,40){\line(1,0){100}}

\put(10,60){\line(-1,1){15}}

\put(00,73){$\mu$}
\put(-6,65){$i$}

\put(160,50){\line(1,0){50}}
\put(160,40){\line(1,0){50}}
\put(160,10){\line(1,0){50}}

\put(119,45){$\ldots\ldots\ldots\ldots$}
\put(119,25){$\ldots\ldots\ldots\ldots$}
\put(119,5){$\ldots\ldots\ldots\ldots$}
\put(0,54){$1$}
\put(0,44){$2$}
\put(-5,4){$e^{NR}$}
\put(219,25){$\ldots\ldots\ldots\ldots$}

\put(10,65){\vector(1,0){50}}
\put(60,65){\vector(-1,0){50}}

\put(60,65){\vector(1,0){50}}
\put(110,65){\vector(-1,0){50}}

\put(160,65){\vector(1,0){50}}
\put(210,65){\vector(-1,0){50}}

\put(30,68){$\gamma N$}
\put(80,68){$\gamma N$}
\put(180,68){$\gamma N$}

\put(12,43){\small $\varphi_N^{(1)}(\tilf_N(v^N)\!=\!2)$}
\put(12,3){\small $\varphi_N^{(1)}(\tilf_N(v^N)\!=\!e^{NR})$}

\put(62,43){\small $\varphi_N^{(2)}(\tilf_N(v^N)\!=\!2)$}
\put(62,3){\small $\varphi_N^{(2)}(\tilf_N(v^N)\!=\!e^{NR})$}

\put(162,43){\small $\varphi_N^{(\mu)}(\tilf_N(v^N)\!=\!2)$}
\put(162,3){\small $\varphi_N^{(\mu)}(\tilf_N(v^N)\!=\!e^{NR})$}

\put(10,50){\textcolor{lightgray}{\rule{200\unitlength}{10\unitlength}}}
\put(12,53){\small $\varphi_N^{(1)}(\tilf_N(v^N)\!=\!1)$}
\put(62,53){\small $\varphi_N^{(2)}(\tilf_N(v^N)\!=\!1)$}
\put(162,53){\small $\varphi_N^{(\mu)}(\tilf_N(v^N)\!=\!1)$}

\put(119,55){$\ldots\ldots\ldots\ldots$}
\put(10,00){\line(0,1){60}}
\put(60,0){\line(0,1){60}}
\put(110,0){\line(0,1){60}}
\put(160,0){\line(0,1){60}}
\put(210,0){\line(0,1){60}}

\put(10,60){\line(1,0){200}}
\put(10,50){\line(1,0){200}}
\put(10,40){\line(1,0){200}}
\put(10,10){\line(1,0){200}}
\end{picture} 
\caption{\Red{Illustration of the random codebook where $R=R(D)+2\varepsilon$ and $\mu\in\bbN$ indexes the different MYICBs. Each row is an infinite-length codeword (cf.~\cite[Eq.~(31)]{Yury2011}), partitioned into blocks of length $\gamma N$ each. Each entry of this matrix which is  of size $\exp(nR)$ by $\infty$ is sampled i.i.d.\ from a CAID $P_X^*$. Note that if $V^N\in\calL_N$, then $\tilf_N(V^N)=1$ so we use the first row of this codebook (shaded). Otherwise if $V^N\notin\calL_N$, then we use the Marton encoder   $\tilf_N$ (given in Lemma \ref{keycor}) to map $V^N$ to one of   $e^{NR}$ rows in the matrix above. }}
\label{fig:random_codebook}
\end{figure*}

\Red{We use  a random coding argument~\cite{Shannon48} to demonstrate the existence of a code satisfying~\eqref{eqachieve}.  Let $\gamma\in (0,1)$ be chosen later (cf.\ \eqref{eqn:defGamma}).  First, the encoder creates a random codebook which consists of $\exp(N(R(D)+2\eps))$ codewords, each  of infinite length but partitioned into blocks of length $\gamma N$ as shown in Fig.~\ref{fig:random_codebook}. Each block\footnote{We assume that $\gamma N$ is an integer; otherwise we regard it as $\lfloor\gamma N\rfloor$.} $\varphi_N^{(\mu)}( i) \sim (P_X^*)^{\gamma N}$ for each $(i,\mu) \in \{1,2,\cdots,\exp(N(R(D)+2\eps))\} \times \bbN$ and $P_X^* \in\argmax_{P_X} I(X;Y)$ is a capacity-achieving input distribution (CAID).\footnote{The idea of using infinite-length codewords for VLF-codes has been used for example in~\cite[Eq.~(31)]{Yury2011}.} The codebook is revealed to the decoder. Given a source sequence $v^N \in \calV^N$ and the random codebook, the random  codewords sent in the message phases of YICBs of length $\gamma N$ (i.e. using the functions $\{\varphi_N^{(\mu)}\}_{\mu\in\bbN}$)~\cite{Gallager1965a} are independent from YICB to YICB.} 
 
Choose a pair of input symbols $(x_0,x_0') \in \calX^2$ such that 
\begin{align}
\label{choice}
(x_0,x_0')=\argmax_{(x,x') \in\calX^2}D(P_{Y|X}(\cdot|x)\,\|\,P_{Y|X}(\cdot|x'))
\end{align} and assign the two (repetition) codewords 
\begin{equation}
\bx_{\rmc}=(x_0,x_0,\cdots,x_0)\quad \mbox{and} \quad \bx_{\rme}=(x_0',x_0',\cdots,x_0') \label{eqn:repet}
\end{equation}
 each of length $(1-\gamma)N$ to control signals $\rmc$ and $\rme$, respectively.\\  

\noindent\underline{Step 1: Encoding and Decoding Strategies}

\Red{We only analyze the case $R(D)<C$ since~\eqref{eqachieve} trivially holds for $R(D)\geq C$. Take a sufficiently small $\eps>0$ such that $R(D)+3\eps<C$. Recall the $(|\calV|^N,N)$-(Marton) block code  $(\tilde{f}_{N},\tilde{g}_{N})$ in Lemma~\ref{keycor}. Let $\rmc,\rme$ be the two control messages in the Yamamoto-Itoh coding scheme. We modify a block of length $N$ of the Yamamoto-Itoh coding scheme~\cite{YamamotoItoh1979}, which consists of two phases (see Fig.~\ref{fig:YI1979}) each of length $\gamma N$ (message phase) and $(1-\gamma)N$ (control phase)    as follows (see Figs.~\ref{fig:message_ph} and~\ref{fig:control_ph}):}
\begin{itemize}[leftmargin=*]
\item \Red{For any $v^N \in \calL_N$, the encoder maps $v^N$ to $\tilde{f}_{N}(v^N)=1$. If $v^N\notin\calL_N$, the encoder  maps $v^N$ to $\tilf_N(v^N)\in  \calM:= \{1,2,\ldots,\exp(N(R(D)+2\eps))\}$. See Fig.~\ref{fig:random_codebook} and its caption.}
\item \Red{Then, the encoder sends $M=\tilde{f}_{N}(v^N) \in \calM$ over the forward link of the DMC, and uses the same two-phase coding block as in a YICB of length $N$.  A {\em modified YICB} or \emph{MYICB} is characterized by the tuple of  functions  $((\varphi_N,\phi_N), (\tilde{\varphi}_N,\tilde{\phi}_N) )$,  where $(\varphi_N,\phi_N)$ is the channel  encoder-decoder pair for the message phase and $(\tilde{\varphi}_N,\tilde{\phi}_N)$ is the channel  encoder-decoder pair in the control phase~\cite{YamamotoItoh1979}; see Figs.~\ref{fig:message_ph} and \ref{fig:control_ph}. Here,    $\varphi_N\in\{\varphi_N^{(\mu)}:\calM\to\calX^{\gamma N}\}_{\mu \in \bbN} $ is the random encoding function  as described in Fig.~\ref{fig:random_codebook}.  The other functions $\phi_N,  \tilde{\varphi}_N,\tilde{\phi}_N$ are deterministic~\cite{YamamotoItoh1979}.  At the end of the message phase, the decoder maps the decoded message $\hatM={\hathh}_{N}(v^N) := \phi_N( Y^{\gamma N} )\in \calM$ to  $\tilde{g}_{N}( {\hathh}_{N}(v^N)) =\tilg_N(\hatM)\in \calV^N$. Here,  $Y^{\gamma N}\in\calY^{\gamma N}$ is distributed according to $P_{Y|X}^{\gamma N}( \fndot \mid  \varphi_N(\tilf_N(v^N)) )$.  Thanks to the noiseless feedback, the encoder can emulate the decoder to estimate $\tilg_N(h_N(v^N))=\tilg_N(\hatM)\in \calV^N$ at the end of the message phase. If  it is ascertained  by the encoder that $d(v^N,\tilg_N(h_N(v^N)))>D$, the encoder sends $\rme$ in the control phase; otherwise it sends $\rmc$ in the control phase. The codewords corresponding to control messages $\rmc$ and $\rme$ form a repetition codebook; cf.\ \eqref{eqn:repet}.}


\begin{figure*}
\centering
\setlength{\unitlength}{.4mm}
\begin{picture}(310,107)
\put(50,00){\line(1,0){50}}
\put(50,30){\line(1,0){50}}
\put(50,00){\line(0,1){30}}
\put(100,00){\line(0,1){30}}

\put(150,00){\line(1,0){50}}
\put(150,30){\line(1,0){50}}
\put(150,00){\line(0,1){30}}
\put(200,00){\line(0,1){30}}

\put(50,90){\line(1,0){50}}
\put(50,120){\line(1,0){50}}
\put(50,90){\line(0,1){30}}
\put(100,90){\line(0,1){30}}

\put(150,90){\line(1,0){50}}
\put(150,120){\line(1,0){50}}
\put(150,90){\line(0,1){30}}
\put(200,90){\line(0,1){30}}

\put(50,00){\line(1,0){50}}
\put(50,30){\line(1,0){50}}
\put(50,00){\line(0,1){30}}
\put(100,00){\line(0,1){30}}

\put(250,45){\line(1,0){50}}
\put(250,75){\line(1,0){50}}
\put(250,45){\line(0,1){30}}
\put(300,45){\line(0,1){30}}

\put(50,15){\vector(-1,0){50}}
\put(150,15){\vector(-1,0){50}}
\put(275,15){\vector(-1,0){75}}
\put(275,15){\line(0,1){30}}

\put(00,105){\vector(1,0){50}}
\put(100,105){\vector(1,0){50}}
\put(200,105){\line(1,0){75}}
\put(275,105){\vector(0,-1){30}}

\put(6,110){$V^N\in\calV^N$}
\put(110,110){$M\in\calM$}
\put(280,90){$X^{\gamma N}\in\calX^{\gamma N}$}
\put(280,30){$Y^{\gamma N}\in\calY^{\gamma N}$}
\put(110,20){$\hatM\in\calM$}
\put(6,20){$\hatV^N\in\calV^N$}

\put(70,103){$\tilf_N$}
\put(170,103){$\varphi_N^{ (\mu) }$}

\put(265,57){$P_{Y|X}^{\gamma N}$}

\put(170,13){$\phi_N$}
\put(70,13){$\tilg_N$}
\end{picture}
\caption{\Red{Block diagram for the coding scheme during the message phase at each repeated MYICB indexed by $\mu\in\bbN$.  Note that $\hatM=h_N(v^N)=\phi_N(Y^{\gamma N})$. The source code $(\tilf_N,\tilg_N)$ is given in Lemma~\ref{lem2}. The channel encoder  $\varphi_N^{(\mu)}$ is described in  Fig.~\ref{fig:random_codebook} and channel decoder  $\phi_N$ is the maximum likelihood decoder~\cite{Gallager1965a} (leading to \eqref{eqn:rcee}).}}
\label{fig:message_ph}
\end{figure*}
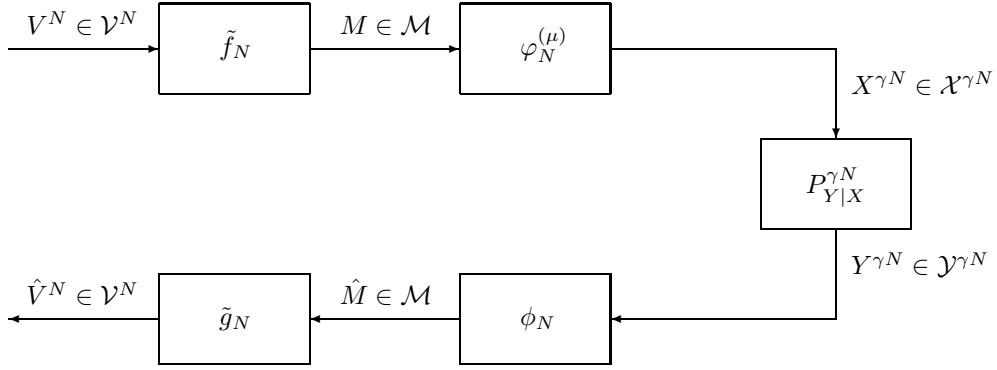

\item  If the encoder knows that the decoder decodes message $\rmc$ in the control phase, it 
resends  a {\em new} source sequence in the message phase of the next MYICB. Otherwise, the encoder sends the {\em same} source sequence as the previous MYICB.
\end{itemize}
\Red{To form a variable-length    code from the MYICB, we repeat  the  transmission of  independent MYICBs of lengths $N$ and also define a stopping time and a final decision. The desired code is created by repeating   MYICBs  at times $n=\mu N+1, \mu \in \bbN$ if the received control signal at time $ \mu N$ is $\rme$. The block diagrams for the coding schemes during the message and control phases at each repeated MYICB indexed by $\mu \in \bbN$ are shown in  Figs.~\ref{fig:message_ph}  and~\ref{fig:control_ph} respectively. } \\

\begin{figure*}
\centering
\setlength{\unitlength}{.4mm}
\begin{picture}(360,50)
\put(50,00){\line(1,0){50}}
\put(50,30){\line(1,0){50}}
\put(50,00){\line(0,1){30}}
\put(100,00){\line(0,1){30}}

\put(00,15){\vector(1,0){50}}

\put(150,00){\line(1,0){50}}
\put(150,30){\line(1,0){50}}
\put(150,00){\line(0,1){30}}
\put(200,00){\line(0,1){30}}

\put(100,15){\vector(1,0){50}}

\put(250,00){\line(1,0){50}}
\put(250,30){\line(1,0){50}}
\put(250,00){\line(0,1){30}}
\put(300,00){\line(0,1){30}}

\put(200,15){\vector(1,0){50}}

\put(300,15){\vector(1,0){50}}

\put(10,20){$\rmc$ or $\rme$}
\put(70,13){$\tilde{\varphi}_N$}

\put(110,20){$\bx_{\rmc}$ or $\bx_{\rme}$}
\put(160,13){$P_{Y|X}^{(1-\gamma)N}$}

\put(210,21){$Y_{ \gamma N+1}^N $}

\put(270,13){$\tilde{\phi}_N$}

\put(310,20){$\rmc$ or $\rme$}
\end{picture}
\caption{\Red{Block diagram for the coding scheme during the control phase. The codewords $\bx_\rme \in\calX^{(1-\gamma)N}$ and $\bx_\rmc\in\calX^{(1-\gamma)N}$ are defined in \eqref{choice} and \eqref{eqn:repet}. }}
\label{fig:control_ph}
\end{figure*}

\noindent\underline{Step 2: Upper Bound on   Excess-Distortion and Retransmission}\\\underline{Probabilities}

Define
\begin{align}
\label{eqn:P1te_v}
\rvP_{1\rme} (v^N)&:= \bbP \big(\hathh_N(V^N) \ne \tilf_N(V^N)\, \big|\,V^N=v^N \big) ,\\
 \label{eqn:P1te}
\rvP_{1\rmT\rme} &:= \bbP \big(\hathh_N(V^N) \ne \tilf_N(V^N)\big) \nn\\*
&=\sum_{ v^N \in \calV^{N} } \rvP_{1\rme} (v^N)P_{V^N}(v^N),\\
\calF_N &:= \{ d(V^N, \tilg_N(h_N(V^N)) )\le D\},\\
\calG_N&:=\{\tilde{\phi}_N (Y_{\gamma N+1}^{N})=\rmc\},\\
\calE_N&:=\{d(V^N,\hatV^N(N))>D\},\\
\label{definece}
\rvP_{\rm{2ce}}&:=\bbP\left(\calG_N^c|\calF_N\right),  \\
\label{defineec}
\rvP_{\rm{2ec}}&:=\bbP\left(\calG_N|\calF_N^c\right),  \\
\label{defineRME}
\rvP_{\rmE}&:=\bbP(\calE_N) ,\\
\label{defineMX}
\rvP_{\rm{RT}}&:=\bbP(\calG_N^c).
\end{align}
Here, $\rvP_{1\rme} (v^N)$ and $\rvP_{1\rmT\rme}$  are respectively the conditional error probability  when sending $v^N \in \calV^N$ and the average error probability over all source sequences $v^N\in\calV^N$ of the message phase (phase 1) of the MYICB. 
In addition, $\rvP_{2\rme\rmc}$ and $\rvP_{2\rmc\rme}$  are respectively  the  probabilities for the transitions $\rme \to \rmc$ and $\rmc \to \rme$  in the control phase (phase 2) of the MYICB. Next, $\rvP_{\rmE}$ is the average excess-distortion probability over all $v^N \in \calV^N$ of the MYICB. 
Finally,  $\rvP_{\rm{RT}}$ is the average retransmission probability over all $v^N \in \calV^N$ of the MYICB.

Now, if $v^N\notin  \calL_N$ and $h_N(v^N) =\tilf_N(v^N)$ we have that $d(v^N, \tilg_N(h_N(v^N)))=d(v^N,\tilg_N(\tilf_N(v^N))) \leq D$ by the definition of $\calL_N$ in~\eqref{eqn:defLN}. Hence, \blue{on the event $\{V^N\notin\calL_N\}$ we have \Red{$\calF_N\supseteq \{ h_N(V^N) =\tilf_N(V^N)\} $}  and by complementation, it follows from the definition of $\rvP_{1\rme} (v^N)$ in~\eqref{eqn:P1te_v} that }
\begin{align}
\label{interfact}
\rvP_{1\rme} (v^N) \geq \bbP(\calF_N^c|V^N=v^N)\quad \forall\; v^N \notin \calL_N,
\end{align} 
Hence, we have
\begin{align}
\bbP(\calF_N^c)
&=\sum_{v^N \notin \calL_N}P_{V^N}(v^N)\bbP(\calF_N^c|V^N=v^N)\nn\\*
&\qquad+\sum_{v^N \in \calL_N}P_{V^N}(v^N)\bbP(\calF_N^c|V^N=v^N)\\
\label{mathbad}
&\leq \sum_{v^N \notin \calL_N}P_{V^N}(v^N)\rvP_{1\rme} (v^N) \!+\!\sum_{v^N \in \calL_N}\!\!P_{V^N}(v^N)\\
&\leq \sum_{v^N \in\calV^n}P_{V^N}(v^N)\rvP_{1\rme} (v^N)\!+\!\sum_{v^N \in \calL_N}\!\! P_{V^N}(v^N)\\
\label{badmath}
&\leq \rvP_{1\rmT\rme} + \bbP(V^N \in \calL_N),
\end{align} where~\eqref{mathbad} follows from~\eqref{interfact} and~\eqref{badmath} follows from~\eqref{eqn:P1te}.

Now our purpose is to bound the excess-distortion probability $\rvP_{\rmE}$ defined in~\eqref{defineRME}. \blue{On the event  $\calF_N$, the encoder sends $\rmc$ in the control phase. If the decoder receives $\rme$ in the control phase, the decoder discards the decoding result $\tilg_N(h_N(V^N))$ in the message phase of this MYICB. In other words, for each MYICB, on the event $\calF_N \cap \calG_N^c$, the decoder does not regard $\tilg_N(h_N(V^N))$ as the final decision, so an excess-distortion event does not occur. 
This means that $\calE_N \cap (\calF_N \cap \calG_N^c)=\emptyset$, so $\calE_N \subset (\calF_N\cap\calG_N^c)^c$. It follows that
\begin{align}
\bbP\big(\calE_N|\calF_N\cap \calG_N^c\big)=0.\label{exp1}
\end{align}
In addition, if  the decoder receives $\rmc$ in the control phase (i.e., $\calG_N$ is true), the decoder adopts the estimate in the message phase (i.e., $\hatV^N(N)=\tilg_N(h_N(V^N))$) as the final decision. This means that $\calE_N \cap (\calF_N \cap \calG_N)=\calF_N^c \cap (\calF_N \cap \calG_N)$. It follows that 
\begin{align}
&\bbP\big(\calE_N|\calF_N\cap\calG_N\big)=0.\label{exp2}
\end{align}}
From~\eqref{exp1} and~\eqref{exp2} and the law of total probability, 
\begin{align}
\bbP(\calE_N|\calF_N)&=\bbP\big(\calE_N|\calF_N\cap\calG_N^c\big)\bbP(\calG_N^c|\calF_N) \nn\\*
&\qquad+\bbP\big(\calE_N|\calF_N\cap\calG_N\big)\bbP(\calG_N|\calF_N)=0.
\label{fact33R2}
\end{align}
\blue{Similarly, on the event  $\calF_N^c$, the encoder sends $\rme$ in the control phase. If the decoder receives $\rme$ in the control phase, the decoder discards the decoding result $\tilg_N(h_N(V^N))$ in the message phase of this MYICB. In other words, for each MYICB, on the event $\calF_N^c \cap \calG_N^c$, the decoder does not regard $\tilg_N(h_N(V^N))$ as the final decision, so an excess-distortion event does not occur. 
This means that $\calE_N \cap (\calF_N^c \cap \calG_N^c)=\emptyset$. Similarly to \eqref{exp1}, it follows that
\begin{align}
\bbP\big(\calE_N|\calF_N^c\cap \calG_N^c\big)=0.\label{exp3}
\end{align}
In addition, if  the decoder receives $\rmc$ in the control phase (i.e., $\calG_N$ is true), the decoder adopts the estimate in the message phase (i.e., $\hatV^N(N)=\tilg_N(h_N(V^N))$) as the final decision. This means that $\calE_N \cap (\calF_N^c \cap \calG_N)=\calF_N^c \cap (\calF_N^c \cap \calG_N)=\calF_N^c \cap \calG_N$. Similarly to \eqref{exp2}, follows that 
\begin{align}
&\bbP\big(\calE_N|\calF_N^c\cap\calG_N\big)=1.\label{exp4}
\end{align}
From~\eqref{exp3} and~\eqref{exp4} and the  law of total probability, }
\begin{align}
\bbP\big(\calE_N|\calF_N^c\big)&=\bbP\big(\calE_N|\calF_N^c\cap\calG_N^c\big)\bbP(\calG_N^c|\calF_N^c) \nn\\*
&\qquad+ \bbP\big(\calE_N|\calF_N^c\cap \calG_N\big)\bbP(\calG_N|\calF_N^c)\\
\label{facto1}
&=\bbP\big(\calG_N|\calF_N^c\big).
\end{align}
Observe that
\begin{align}
\rvP_{\rmE} &=\bbP(\calE_N|\calF_N^c)\bbP(\calF_N^c)+\bbP(\calE_N|\calF_N)\bbP(\calF_N)\\
\label{eqrevised:1}
&=\bbP(\calE_N|\calF_N^c)\bbP(\calF_N^c)\\
\label{eqrevised:2}
&=\bbP(\calG_N|\calF_N^c)\bbP(\calF_N^c)\\
\label{eqrevised:3}
&=\rvP_{2\rme\rmc}\bbP(\calF_N^c)\\
&\leq \rvP_{2\rme\rmc},  \label{eqn:P2ce_bd}
\end{align}
where~\eqref{eqrevised:1} follows from~\eqref{fact33R2},~\eqref{eqrevised:2} follows from~\eqref{facto1}, and~\eqref{eqrevised:3} follows from the definition of $\rvP_{2\rme\rmc}$ in~\eqref{defineec}. 

On the other hand, for the MYICB (as described in Step~1 above), $\rvP_{\rm{RT}}$, defined in \eqref{defineMX}, can  bounded as follows
\begin{align}
\label{test2000}
\rvP_{\rm{RT}}&=\bbP(\calF_N^c)  \bbP\left(\calG_N^c|\calF_N^c\right) +  \bbP(\calF_N) \bbP\left(\calG_N^c|\calF_N\right)\\
\label{eq47facto}
&=\bbP(\calF_N^c) (1-\rvP_{2\rme\rmc})+ \bbP(\calF_N)\rvP_{2\rmc\rme}\\
\label{eqn:upper_bd_PX}
&\leq \big[\rvP_{1\rmT\rme}+\bbP\big(V^N \in \calL_N\big)\big](1-\rvP_{2\rme\rmc})+\rvP_{2\rmc\rme}, 
\end{align} where~\eqref{test2000} follows from~\eqref{defineMX} and total law of probability, and~\eqref{eq47facto} follows from~\eqref{definece} and~\eqref{defineec}, and~\eqref{eqn:upper_bd_PX} follows from~\eqref{badmath}.

Now since $|\calR(\tilde{f}_{N})|=|\calM|= \exp(N(R(D)+2\eps))$, by Gallager's~\cite{Gallager1965a} error exponent analysis for maximum likelihood decoding\footnote{The random coding bound holds for any distribution of messages  as it holds under the maximum probability of error formalism.} we know that if 
\begin{align}
\label{condl1}
\frac{R(D)+2\eps}{\gamma}<C,
\end{align} then\footnote{We use the notation $a_n\dotleq b_n$ to mean $\limsup_{n\to\infty}\frac{1}{n}\log\frac{a_n}{b_n}\le 0$.}
\begin{align}
\rvP_{1\rmT\rme} \dotleq \exp\big(-\gamma N \tilde{F}(R(D),\eps,\gamma)\big), \label{eqn:rcee}
\end{align}
where   $\tilde{F}(R(D),\eps,\gamma)>0$ is the random coding error exponent~\cite{Gallager1965a}. \Red{Note that $\varphi_N^{(\mu)} \sim (P_X^*)^{\gamma N}$ for all $\mu \in \bbZ^+$.}

Besides,  analyzing  the same decoding strategy for the control mode similarly to that in~\cite{YamamotoItoh1979}, we have
\begin{align}
\label{condl2}
\rvP_{2\rmc\rme} \dotleq \exp\left(-N \beta \right),
\end{align} for some $\beta>0$.

In addition, by Chernoff-Stein's lemma~\cite[Chapter~11]{Cov06} and the choice of $(x_0,x_0')$ in~\eqref{choice}, we also have
\begin{align}
\lim_{N\to \infty} -\frac{\ln \rvP_{2\rme\rmc}}{(1-\gamma) N}=B.
\end{align}
It follows from~\eqref{eqn:P2ce_bd} that
\begin{align}
\liminf_{N\to \infty} -\frac{\ln \rvP_{\rmE}}{N}&\geq (1-\gamma)\liminf_{N\to \infty} -\frac{\ln \rvP_{2\rme\rmc}}{(1-\gamma) N}\\
\label{result1}
&=(1-\gamma)B.
\end{align}
By choosing 
\begin{align}
\gamma=\frac{R(D)+3\eps}{C} \in (0,1), \label{eqn:defGamma}
\end{align} to satisfy~\eqref{condl1}, we have from~\eqref{result1} that
\begin{align}
\liminf_{N\to \infty} -\frac{\ln \rvP_{\rmE}}{N} \geq B\left(1-\frac{R(D)+3\eps}{C}\right). \label{eqn:exp_PE}
\end{align}
Note that with this choice of $\gamma$, from \eqref{eqn:upper_bd_PX}, \eqref{zeroten}, \eqref{eqn:rcee}, and \eqref{condl2}, we have
\begin{align}
\label{eq44mod}
\rvP_{\rm{RT}} &\leq \rvP_{1\rmT\rme}+\bbP\big(V^N \in  \calL_N\big)+ \rvP_{2\rmc\rme}\\
\label{eq45new}
& \dotleq \exp(-\alpha N),
\end{align} where $\alpha:=\min\{\beta,\gamma\tilde{F}(R(D),\eps,\gamma), F(P_V,R(D)+\eps,D)/2\}>0$.  \\

\noindent\underline{Step 3: Repetition of Independent MYICBs to Form a}\\\underline{Variable-Length Code}

\Red{Note that $\{Y_{(t-1)N+\gamma N+1}^{(t-1)N+N}:t\in\bbN\}$ are random vectors which are mutually independent of each other  by the  generation of the random codebook in Fig.~\ref{fig:random_codebook} (each element is i.i.d.\ generated according to $P_X^*$).} 

 The  stopping time $\tau_N$ can be defined as follows:
\begin{enumerate}
\item If $n=\mu N, \mu \in\bbN\setminus\{1\}$, we define
\begin{align}
\label{eq47new}
\bone\{\tau_N=n\}& = \prod_{t=1}^{\mu-1} \bone\left\{\tilde{\phi}_N\big(Y_{(t-1)N+\gamma N+1}^{(t-1)N+N}\big)=\rme \right\}   \nn\\*
&\quad \times\bone\left\{\tilde{\phi}_N \big(Y_{(\mu-1)N+\gamma N+1}^{n}\big)=\rmc\right\},
\end{align} where $\tilde{\phi}_N$ is the decoder of MYICB for the control mode defined in Step~1 above.
\item If $n = N$, we define 
\begin{align}
\bone\{\tau_N=n\}=\bone \left\{\tilde{\phi}_N\big(Y_{\gamma N+1}^N\big)=\rmc\right\};
\end{align}
\item Otherwise,
\begin{align}
\bone\{\tau_N=n\}=\bone\{\emptyset\}.
\end{align}
\end{enumerate}

In addition, the estimated sequence of the transmitted sequence at the stopping time $\tau_N$ is 
\begin{align}
\hat{V}^N(\tau_N):=\tilde{g}_{N}\Big(\phi_N \big(Y_{\tau_N-N}^{\tau_N- (1-\gamma)N }\big)\Big).
\end{align}
Note that $\phi_N \big(Y_{\tau_N-N}^{\tau_N- (1-\gamma)N }\big)$ is the estimated message corresponding to $\tilde{f}_N$ at the stopping time $\tau_N$. \\

\noindent\underline{Step 4: Bounding the Expectation of the Stopping Time}

 By the definition of $\tau_N$ in Step 3, we have $\{\tau_N=n\}\in \sigma(Y^n)$ for all $ n\in\bbN$.  Furthermore, by the proposed  transmission scheme, \Red{we deduce that $\{Y_{(t-1)N+1}^{(t-1)N+N}: t\in\bbN\}$ are i.i.d.\ random vectors (from the construction of the variable-length code in Step 3 above) since we use the random function $\varphi_N$ (which generates independent coding blocks)  for the source sequence $V^N$ at each repeated MYICB at times $n=\mu N+1$ for $\mu\in\bbN$ (cf.\ Fig.~\ref{fig:random_codebook}). Therefore, we have that  }
\begin{align}
\label{eq51newbest}
\bbP\left(\tilde{\phi}_N\big(Y_{(t-1)N+\gamma N+1}^{(t-1)N+N}\big)=\rme\right) &= \bbP\left(\tilde{\phi}_N \big(Y_{\gamma N+1}^N\big)=\rme\right)\nn\\*
&=\rvP_{\rm{RT}}   \quad \forall\, t\in\bbN,
\end{align}
where the last equality follows from the definition of $\rvP_{\rm{RT}}   $ in \eqref{defineMX}. 
It follows from~\eqref{eq47new} and our retransmission scheme that 
\begin{align}
\label{eqtestrev}
\bbP(\tau_N=n)&=\begin{cases} \big(\rvP_{\rm{RT}} \big)^{\frac{n}{N}-1}  \big(1-\rvP_{\rm{RT}} \big),&\mbox{if}\;N\mid n\\ 0,&\mbox{otherwise}\end{cases}.
\end{align}
Observe that
\begin{align}
\sum_{n=0}^{\infty} \bbP(\tau_N=n) &= \sum_{\mu=1}^{\infty} \big(\rvP_{\rm{RT}} \big)^{\mu-1}  \big(1-\rvP_{\rm{RT}} \big)
\label{newr2}
= 1,
\end{align} where~\eqref{newr2} follows from~\eqref{eqtestrev}. Thus, from~\eqref{eq47new} and~\eqref{newr2}, we deduce that  $\tau_N$ is indeed a stopping time with respect to the filtration $\{\sigma(Y^n)\}_{n=0}^{\infty}$.

Moreover,  we also have
\begin{align}
\bbE(\tau_N)&=\sum_{n=0}^{\infty} n \bbP(\tau_N=n)\\
\label{eq45mod2}
&=\sum_{\mu=1}^{\infty} \mu N \big(\rvP_{\rm{RT}} \big)^{\mu-1}  \big(1-\rvP_{\rm{RT}} \big) \\
\label{eq48mod2}
&= \frac{N}{ 1- \rvP_{\rm{RT}} }\\
&\le N+o(1),\label{eq58new}
\end{align}
where~\eqref{eq45mod2} follows from~\eqref{eqtestrev}, and~\eqref{eq58new} follows from~\eqref{eq45new}. \\

\noindent\underline{Step 5: Bounding the Excess-Distortion Probability}

Note that 
\begin{align}
&\bbP\left(\{d(\hat{V}^N(\tau_N),V^N)>D\}\cap\{\tau_N=\mu N\}\right) \nonumber\\
&=\bbP\bigg(\bigcap_{t=1}^{\mu-1} \{\tilde{\phi}_N\big(Y_{(t-1)N+\gamma N+1}^{(t-1)N+N}\big)=\rme\} \nn\\*
&\qquad\qquad\cap \{\tilde{\phi}_N\big(Y_{(\mu-1)N+\gamma N+1}^{(\mu-1)N+N}\big)=\rmc\} \nn\\*
&\qquad\qquad\cap \{d(\hat{V}^N(\mu N),V^N)>D\}\bigg) \label{eqn:ecD}\\
&\leq \bbP\bigg(\bigcap_{t=1}^{\mu-1} \{\tilde{\phi}_N\big(Y_{(t-1)N+\gamma N+1}^{(t-1)N+N}\big)=\rme\}  \nn\\*
&\qquad\qquad\cap \{d(\hat{V}^N(\mu N),V^N)>D\}\bigg) \\
&\leq \prod_{t=1}^{\mu-1}\bbP\left(\tilde{\phi}_N\big(Y_{(t-1)N+\gamma N+1}^{(t-1)N+N}\big)=\rme\right)\nn\\*
&\qquad\qquad\times \bbP \big(d(\hat{V}^N(\mu N),V^N)>D\big)\label{eqn:ecE}\\
\label{fivefact}
&=(\rvP_{\rm{RT}})^{\mu-1} \rvP_{\rmE},
\end{align} \blue{where~\eqref{eqn:ecE} follows from the fact that since $V^N$ is independent of $Y_{1}^{(\mu-1)N}$ (because a fresh or independent source sequence is transmitted in each MYICB), both $V^N$ and $\hatV^N(\mu N)$ are independent of $Y_{1}^{(\mu-1)N}$ (but $\hatV^N(\mu N)$ depends on $Y_{(\mu-1)N+1}^{\mu N}$), and~\eqref{fivefact} follows from the definitions of $\rvP_{\rmE}$ in~\eqref{defineRME} and $\bbP\big(\tilde{\phi}_N\big(Y_{(t-1)N+\gamma N+1}^{(t-1)N+N}\big)=\rme\big)$ in~\eqref{eq51newbest}. }
Using the above calculation, now observe that
\begin{align}
\label{eq58R2}
&\bbP\big(d(\hatV^N(\tau_N),V^N)>D \big)\nn\\*
&=\sum_{n=0}^{\infty} \bbP\big(\{d(\hat{V}^N(\tau_N),V^N)>D\}\cap\{\tau_N=n\}) \\
\label{eq61modR2}
&=\sum_{\mu=1}^{\infty} \bbP\big(\{d(\hat{V}^N(\tau_N),V^N)>D\}\cap\{\tau_N=\mu N\}) \\
\label{eq62modR2}
&\leq  \sum_{\mu=1}^{\infty} (\rvP_{\rm{RT}})^{\mu-1} \rvP_{\rmE}\\
 \label{fact3}
&= \frac{\rvP_{\rmE}}{1-\rvP_{\rm{RT}}},
\end{align} where~\eqref{eq61modR2} follows from~\eqref{eqtestrev} and \eqref{eq62modR2} follows from \eqref{fivefact}. 

Therefore, the resultant excess-distortion exponent   is
\begin{align}
&\liminf_{N\to \infty}-\frac{\ln \rvP_{\rmd}(N,D)}{N}\nn\\*
 &\ge \liminf_{N\to \infty}  \left\{-\frac{\ln \rvP_{\rmE}}{N} +\frac{\ln(1-\rvP_{\rm{RT}})}{N} \right\}\label{eqn:PdPEPX} \\
&\geq B\left(1-\frac{R(D)+3\eps}{C}\right)+ \lim_{N\to \infty} \frac{\ln(1-\exp(-\alpha N))}{N} \label{eqn:use_exp_PE}\\
&=B\left(1-\frac{R(D)+3\eps}{C}\right),
\end{align}
where \eqref{eqn:PdPEPX} follows from~\eqref{errordef} and \eqref{fact3}, and \eqref{eqn:use_exp_PE} follows from \eqref{eqn:exp_PE} and \eqref{eq45new}. 
Since $\eps>0$ can be made arbitrarily small, 
 this concludes the proof of~\red{Proposition~\ref{achieve:proof}}.
\end{IEEEproof}
\section{Converse Proof}\label{sec:converse}
\red{The converse proof techniques are partly based on~\cite{Berlin2009a}. There are some changes in proof techniques in various lemmata to account for the fact that we are dealing with lossy joint-source channel coding (instead of the much simpler channel coding). Two main novel techniques are developed in this paper. They  include (i) a distortion-MAP decoding rule which replaces for the MAP decoding rule in the proof of~\cite[Lemma 1]{Berlin2009a} and (ii) developing somewhat non-standard data processing inequalities of divergences, replacing the use of Fano's inequality also in the proof of~\cite[Lemma 1]{Berlin2009a}. The converse proof is in parallel to the two-phase modified Yamamoto-Itoh coding scheme developed in our achievability proof in Section~\ref{sec:ach}.} 

Fix a $(|\calV|^N, N)$-variable-length joint source-channel code with feedback in as Definition~\ref{def1}. This specifies the excess-distortion probability $\rvP_{\rmd}(N,D)$. Define the posterior distribution $P_{V^N|Y^n}(v^N|y^n)$ as
\begin{align}
&P_{V^N|Y^n}(v^N|y^n)\nn\\*
&:=\frac{\prod_{k=1}^n P_{Y|X}(y_k|f_k(v^N,y^{k-1})) \prod_{j=1}^NP_V (v_j)}{\sum_{v^N} \prod_{k=1}^n P_{Y|X}(y_k|f_k(v^N,y^{k-1})) \prod_{j=1}^NP_V (v_j)}.
\end{align}
Define the random stopping times
\begin{align}
\label{definetau^*}
\tau^*_N&:=\inf\bigg\{n: \min_{v^N \in V^N}\sum_{w^N \in \calV^N\setminus \calS_D(v^N)} P_{V^N|Y^n}(w^N|Y^n) \nn\\*
&\qquad\qquad\qquad\qquad\leq \delta_N\bigg\},\\
\label{definetau'}
\tau_N'&:=\inf\bigg\{n:  \min_{v^N \in V^N}\sum_{w^N \in \calV^N\setminus \calS_D(v^N)} P_{V^N|Y^n}(w^N|Y^n) \nn\\*
&\qquad\qquad\qquad\qquad\leq \rvP_{\rmd}(N,D)\bigg\},
\end{align} for some sequence $\delta_N \geq \rvP_{\rmd}(N,D)$ to be determined later. \red{The stopping times $\tau^*_N$ and $\tau_N'$ emulate the stopping times of two   phases in the proof of the direct part, i.e.,  the message and the control phases in the modified YICB (presented in Section~\ref{sec:ach}).}  

\red{Next, we state and prove a key lemma concerning the MAP decoding rule. This  rule is inspired by  \cite{Berlin2009a}. In \cite{Berlin2009a}, it was the crux to ensure that the converse proof provided of Burnashev's exponent can be simplified compared to the original converse proof in~\cite{Burnashev1976} involving martingales.}
\begin{lemma} \label{key} For any fixed sequence of encoders $\{f_n\}_{n=1}^{\infty}$ of a $(|\calV|^N,N)$-variable-length joint source-channel code with feedback (Definition~\ref{def1}),  given any fixed blocklength $\tau_N=n\in \bbN$ \blue{and the channel output $Y^n$},  define the distortion-MAP decoding rule as follows
\begin{align}
\hatV^N(n):=\argmax_{v^N \in \calV^N} \sum_{w^N \in \calS_D(v^N)}P_{V^n|Y^n}(w^n|Y^n) \label{decodingR2}.
\end{align}
Then, the distortion-MAP decoding rule achieves the lowest excess-distortion probability (highest excess-distortion exponent) among all decoding rules. In addition, the following holds:
\begin{align}
\label{keyfact}
&\bbP(d(\hat{V}^N(n), V^N)>D|Y^n)\nn\\*
&=\min_{v^N \in \calV^N}\sum_{w^N \in \calV^N\setminus \calS_D(v^N)} P_{V^N|Y^n}(w^N|Y^n)\quad \mbox{a.s.}
\end{align}
\end{lemma}
\begin{IEEEproof} Observe that for each $n\in \bbN$, we have
\begin{align}
&\bbP(d(\hat{V}^N(n), V^N)\leq D)\nn\\*
&=\sum_{y^n\in \calY^n} P_{Y^n}(y^n) \sum_{w^N \in \calS_D(g_n(y^n))}  P_{V^N|Y^n} (w^N|y^n) \\
\label{eqtrick2}
&\leq \sum_{y^n\in \calY^n}P_{Y^n}(y^n) \max_{v^N \in \calV^N} \sum_{w^N \in \calS_D(v^N)}  P_{V^N|Y^n} (w^N|y^n).
\end{align}
Here,~\eqref{eqtrick2} follows from the fact that $g_n(y^n)\in \calV^N$. One important note is that the inequality~\eqref{eqtrick2} becomes an equality if we choose the decoding function $g_n(y^n)=v_0^N$, where $v_0^N$ is in the set 
\begin{align}
&\Bigg\{v_0^N \in \calV^N:\sum_{w^N \in \calS_D(v_0^N)} P_{V^N|Y^n}(w^n|Y^n)\nn\\*
&\qquad=\max_{v^N \in V^N}\sum_{w^N \in \calS_D(v^N)} P_{V^N|Y^n}(w^n|Y^n)\Bigg\}.
\end{align}
Using the distortion-MAP decoding rule, it is easy to see from~\eqref{eqtrick2} that
\begin{align}
&\bbP(d(\hat{V}^N(n), V^N)>D|Y^n)\nn\\*
&= 1- \max_{v^N \in V^N}\sum_{w^N \in \calS_D(v^N)} P_{V^N|Y^n}(w^n|Y^n)\\
&=\min_{v^N \in V^N}\sum_{w^N \in \calV^N\setminus \calS_D(v^N)} P_{V^N|Y^n}(w^n|Y^n).
\end{align}
This means that joint source-channel codes employing the distortion-MAP decoding rule have the smallest    excess-distortion probability if we use the same sequence of encoders $\{f_n\}_{n=1}^{\infty}$ and the same stopping time rule $\tau_N$. This concludes the proof of Lemma~\ref{key}.
\end{IEEEproof}
 From Lemma~\ref{key}, for the purpose of finding upper bound on the distortion reliability function of $(|\calV|^N,N)$-variable-length joint source-channel codes with feedback, it is sufficient to consider codes that use the distortion-MAP decoding rule, and hence~\eqref{keyfact} can be assumed. 
 
\begin{lemma}\label{stopcomp} For all $(|\calV|^N,N)$-variable-length joint source-channel codes with feedback, the following statements hold:
\begin{align}
\label{hmfash}
\bbP(d(\hat{V}^N(\tau_N^*), V^N)>D|Y^{\tau_N^*}) &\leq \delta_N, \quad \mbox{a.s.}\\
\label{compare}
\bbP(d(\hat{V}^N(\tau_N^*), V^N)>D) &\leq \delta_N,\\
\label{eqcompare}
\bbP(d(\hat{V}^N(\tau_N'),V^N)>D) &\leq \rvP_{\rmd}(N,D),\\
\label{finding}
\bbE(\tau_N^*) \leq \bbE(\tau_N')& \leq \bbE(\tau_N).
\end{align} 
\end{lemma}
\begin{IEEEproof} We know from~\eqref{definetau^*} and~\eqref{keyfact} that for all $k\leq \tau_N^*-1$,
\begin{align}
&\bbP(d(\hat{V}^N(k),V^N)>D)\nn\\*
&=\bbE\left[\bbP\big(d(\hat{V}^N(k),V^N)>D\big|Y^k\big) \right]\\
&=\bbE\Big[\min_{ v^N \in \calV^N}\sum_{w^N \in \calV^N\setminus \calS_D(v^N)} P_{V^N|Y^k}(w^N|Y^k) \Big]\\
&> \delta_N.
\end{align}
Moreover, from~\eqref{definetau^*} and~\eqref{keyfact} we also know that 
\begin{align}
&\bbP\big(d(\hat{V}^N(\tau_N^*),V^N)>D \big|Y^{\tau_N^*}\big)\nn\\*
&= \min_{v^N \in \calV^N}\sum_{w^N \in \calV^N\setminus \calS_D(v^N)} P_{V^N|Y^{\tau_N^*}}(w^N|Y^{\tau_N^*}) \\
\label{builtest}
&\leq \delta_N,\enspace \mbox{a.s.},
\end{align} so~\eqref{hmfash} follows from~\eqref{builtest}. The bound in~\eqref{compare} follows by taking expectations on both sides of~\eqref{hmfash}. 

Similarly, from the definition of $\tau_N'$,  we also have~\eqref{eqcompare}. Furthermore, for all $k\leq \tau_N'-1$ 
\begin{align}
\label{cure1}
\bbP(d(\hat{V}^N(k),V^N)>D) > \rvP_{\rmd}(N,D).
\end{align}
Since $\bbP(d(\hat{V}^N(\tau_N), V^N)>D) \leq \rvP_{\rmd}(N,D)$, from~\eqref{cure1} we  have $\tau_N\geq \tau_N'$. In addition, from~\eqref{definetau^*} and~\eqref{definetau'} we obtain $\tau_N' \geq \tau_N^*$ since $\delta_N \geq \rvP_{\rmd}(N,D)$. Hence, we obtain~\eqref{finding}. This concludes the proof of Lemma~\ref{stopcomp}.
\end{IEEEproof}

\begin{lemma}\label{lem2} We have that for $N \to \infty$, 
\begin{align}
\label{important1}
\bbE[\tau_N^*] C \geq  (1-\delta_N) N R(D)+O(\sqrt{N}).
\end{align}
\end{lemma}
\blue{This lemma provides a lower bound on the expectation of the length of the message phase in the direct part of Burnashev's coding scheme~\cite{Burnashev1976}.  } 
\begin{IEEEproof}
Let $\rmQ^{-1}(\cdot)$ be the inverse of the complementary cumulative distribution function of a standard Gaussian. Since $\tau_N^*$ is a stopping time of the joint source-channel coding scheme, from~\eqref{compare} and~\cite[Theorem 5]{kost17} we have
\begin{align}
&\bbE[\tau_N^*] C \nn\\*
&\geq (1-\bbP(d(\hat{V}^N(\tau_N^*), V^N)>D)) N R(D)\nn\\*
&\quad -\sqrt{\frac{N \nu(D)}{2\pi}}\exp\Bigg(\!-\! \frac{\big[\rmQ^{-1}(\bbP\big(d(\hat{V}^N(\tau_N^*), V^N)\!>\! D)\big)\big]^2}{2}\Bigg)\nn\\*
&\quad +O(\ln N)\label{eqn:drop_exp0}\\
&\ge (1-\delta_N) N R(D)+O(\sqrt{N}). \label{eqn:drop_exp}
\end{align} Note that  $\nu(D)$, which is immaterial for the discussions to follows, is defined in~\cite[Eq.~(60)]{kost17}. The bound in \eqref{eqn:drop_exp} holds because the exponential term in \eqref{eqn:drop_exp0} is upper bounded by $1$, $\nu(D)$ is finite, \red{and that~\eqref{compare} holds.}
\end{IEEEproof}
\begin{lemma} \label{Berlinlem} 
For any $(|\calV|^N,N)$ variable-length joint source-channel code with feedback, for $N$ sufficiently large  and if $\lambda \delta_N \geq  \rvP_{\rmd}(N,D)$, the following holds
\begin{align}
\label{eq84best}
\bbE[\tau_N'-\tau_N^*] \geq -\frac{\ln \rvP_{\rmd}(N,D)}{B}+\frac{ \ln\left[\min\left\{\lambda \delta_N,1-\delta_N\right\}\right]-2}{B}.
\end{align} 
\end{lemma} 
\blue{This lemma provides a lower bound for the expectation of the length of the control phase in the direct part. However, note that this control phase does not coincide with the control phase of the MYICB (as discussed in Section~\ref{sec:ach}). It in fact emulates the variable-length control phase in Burnashev's coding scheme~\cite{Burnashev1976}. The expected length of this phase is 
 that of the sequential binary hypothesis test between $\rvH_0:V^N \in \calS_D(\hatV^N(\tau_N^*))$ and $\rvH_1:V^N \notin \calS_D(\hatV^{N}(\tau_N^*)$).} 
\begin{IEEEproof}
Preliminaries for the proof are provided  in Appendix \ref{proof:prep}. The main body of the  proof is in Appendix~\ref{proof:Berlinlem}.
\end{IEEEproof}
   Some remarks and novelties contained in  our proof technique are discussed below.
\begin{itemize}[leftmargin=*]
\item   The proof technique in Berlin {\em et al.}~\cite[Lemma 1]{Berlin2009a} is based on    Fano's technique just as in Burnashev~\cite{Burnashev1976}. \blue{However, for lossy joint-source channel coding,   a na\"ive application of Fano's inequality does not work. To ameliorate the stumbling blocks, we modify several inequalities  in \cite{Berlin2009a} to ensure that the proof idea in~\cite[Lemma 1]{Berlin2009a} works for joint source-channel coding. For example, we replace the inequalities~\cite[Eq.~(12)]{Berlin2009a} and~\cite[Eq.~(13)]{Berlin2009a} by~\eqref{newsun1} in Lemma~\ref{newsun1lem}. To do so necessitates   the  use some data-processing inequalities in non-standard ways. For example, the log-sum inequality is used {\em twice} to bound a certain term (namely $L_{n,m}^{(1)}\ln (L_{n,m}^{(1)}/L_{n,m}^{(2)}) + \overline{L}_{n,m}^{(1)}\ln(\overline{L}_{n,m}^{(1)}/\overline{L}_{n,m}^{(2)})$ where the constituent terms are defined in  \eqref{eq98most1}--\eqref{eq99most}).  Specifically, we show that this term is effectively a nested sum and apply the log-sum inequality to {\em both} inner and outer sums  in~\eqref{tricky1} and~\eqref{eqnew2017} to obtain~\eqref{eqn:lsi} and~\eqref{eqn:msi}. 
 Furthermore, for the lossy source-channel coding problem at hand, we need work with an abstract distortion measure $d:\calV\times\calV\to [0,+\infty)$ rather than directly compare the transmitted  and decoded messages as in the standard channel coding problem studied by Burnashev~\cite{Burnashev1976}.  In particular, we perform a detailed analysis of the distortion-MAP decoding rule in Appendix~\ref{proof:Berlinlem}. }
\item  In addition, we also note that the proof technique in~\cite[Lemma 1]{Berlin2009a} is based on a probabilistic model on a certain observation tree, hence one must assume that the depth of the observation tree is finite, i.e. $T$ is finite in~\cite[Eq.~(14)]{Berlin2009a}. This assumption is essentially correct since $\bbP(T<\infty)=1$ but in general there is a difference between $\bbP(T<\infty)=1$ and $T$ is (almost surely) bounded in stopping time theory. To streamline  our proofs, we use a slightly different way of defining stopping times compared to~\cite[Eq.~(14)]{Berlin2009a}.  Specifically, our stopping times are defined in~\eqref{definetau^*} and~\eqref{definetau'}. Furthermore, some new proof techniques to analyze two stopping times that are not almost surely bounded but only having finite expectations are also given in the proof of Lemma \ref{Berlinlem} contained in Appendices~\ref{proof:prep} and~\ref{proof:Berlinlem}. 
\end{itemize}
\begin{lemma}~\cite[Proposition 2]{Berlin2009a} \label{notepoint} If $B<\infty$, then  for all $v^N\in\calV^N$ and $y^n\in\calY^n$,
\begin{align}
P_{V^N|Y^n}(v^N|y^n) \geq \lambda P_{V^N|Y^{n-1}}(v^N|y^{n-1}),
\end{align}
\red{for some $0<\lambda:=\min_{(x,y) \in \calX \times \calY} P_{Y|X}(y|x)\leq \frac{1}{2}.$}
\end{lemma}
\begin{proposition}[Converse]\label{converse} The following inequality holds
\begin{align}
\label{pro1}
E^*(D)\leq \max\left\{0, B\left(1-\frac{R(D)}{C}\right)\right\}.
\end{align}
\end{proposition}
\begin{IEEEproof}[Proof of Proposition~\ref{converse}] Define
\begin{align}
\label{trivicond}
\beta:=\liminf_{N\to \infty} \frac{-\ln \rvP_{\rmd}(N,D)}{N}.
\end{align} We will consider two cases $\beta>0$ and $\beta=0$. For the case $\beta=0$, we immediately have that $E(D)=0$. Therefore, we only need to consider the case $\beta > 0$.  Define 
\begin{align}
\label{pointnew}
\xi:=\limsup_{N\to \infty} \frac{-\ln \rvP_{\rmd}(N,D)}{N} \geq \beta>0.
\end{align}
 Observe that from~\eqref{trivicond} and~\eqref{pointnew} we also have that for $N$ sufficiently large, 
\begin{align}
\label{newpoint}
 e^{-2N\xi}< \rvP_{\rmd}(N,D) < e^{-\beta N/2}.
\end{align} 
Now, by choosing $\lambda\delta_N := 1/(-\ln\rvP_{\rmd}(N,D)) \geq \rvP_{\rmd}(N,D)$, from the upper bound in~\eqref{newpoint}, we have  that 
\begin{equation}
\lim_{N\to\infty}\delta_N =0. \label{eqn:delta_inf_many}
\end{equation}
On the other hand, observe that
\begin{align}
0\geq \limsup_{N\to \infty} \frac{\ln\left(\lambda \delta_N\right)}{N}&\geq \liminf_{N\to \infty} \frac{\ln\left(\lambda \delta_N\right)}{N}\\
&=\liminf_{N\to \infty}-\frac{\ln(-\ln \rvP_{\rmd}(N,D))}{N}\\
\label{eq94note}
&\ge \liminf_{N\to \infty}-\frac{\ln(2\xi N)}{N}\\
\label{eq95note}
&=0,
\end{align}
where~\eqref{eq94note} follows from the lower bound in~\eqref{newpoint}  and~\eqref{eq95note} follows from the assumption that $\xi>0$ [cf.~\eqref{pointnew}]. Therefore, 
\begin{align}
\label{test1}
\lim_{N\to \infty} \frac{\ln\left(\lambda \delta_N\right)}{N}&=0.
\end{align}
It follows from Lemmas~\ref{stopcomp},~\ref{lem2}, and~\ref{Berlinlem} that for any $(|\calV|^N,N)$-variable-length joint source-channel code, 
\begin{align}
\label{newly}
\bbE(\tau_N) &\geq \bbE[\tau_N']=\bbE[\tau_N^*]+\bbE[\tau_N'-\tau_N^*]\\
\label{suppl}
& \geq \frac{(1-\delta_N) N R(D)+O(\sqrt{N})}{C} \nn\\*
&\qquad -\frac{\ln \rvP_{\rmd}(N,D)}{B}+\frac{ \ln\left(\lambda \delta_N\right)-2}{B}
\end{align}
Hence, we obtain 
\begin{align}
 \liminf_{N\to\infty} \frac{-\ln \rvP_{\rmd}(N,D)}{N} 
&\le B\left(1-\frac{R(D)}{C}\right) \label{eqn:final_conve}
\end{align}
where~\eqref{eqn:final_conve} follows from~\eqref{eq8def},~\eqref{eqn:delta_inf_many},~\eqref{test1}, and \eqref{suppl}. This implies that $E^*(D)$ is upper bounded by the right-hand-side of~\eqref{eqn:final_conve}. This concludes the proof of Proposition~\ref{converse}. 
\end{IEEEproof}
\red{\section{Conclusion} \label{conclu}
The reliability function (optimum error exponent) for the transmission of DMSes  across DMCs  using variable-length lossy source-channel codes with feedback is proved to be $\max\{0, B(1-R(D)/C)\}$. In this setting and in this asymptotic regime, separate source-channel coding is surprisingly optimal. For future work, we note that Draper and Sahai~\cite{Draper2008} demonstrated that as the desired rate of communication approaches the capacity of the forward channel, Burnashev's reliability function~\cite{Burnashev1976} is achievable given any positive-capacity noisy feedback channel. An interesting future work is to find an achievablity coding scheme which achieves our optimal distortion exponent under the same condition as in Draper and Sahai's work~\cite{Draper2008}.
}
\appendices 
\section{Preliminaries for the Proof of Lemma \ref{Berlinlem}}\label{proof:prep}
In all proofs of this section, we use the following    notations for simplicity of presentation:
\begin{align}
\overline{\calS}_D(\hat{V}^N(m))& :=\calV^N\setminus \calS_D(\hat{V}^N(m)),\\
\calK_m&:=\calS_D(\hat{V}^N(m))\times \overline{\calS}_D(\hat{V}^N(m)), \label{eqn:defKm}
\end{align}
and for each given $Y^m=y^m$, define 
\begin{align}
\label{defineQ1}
\calQ_n(\hatV^N(m),y^m)&:=\{y_{m+1}^n\! \in\! \calY_{m+1}^n\!:\! g_n(y^n) \!\in\! \calS_D(\hatV^N(m))\},\\
\label{defineQ2}
\overline{\calQ}_n(\hatV^N(m),y^m)&:= \calY^{n-m}\setminus \calQ_n(\hatV^N(m),y^m).
\end{align}
Note that $\calQ_n(\hatV^N(m),y^m)$ and ${\cal\barQ}_n(\hatV^N(m),y^m)$ are deterministic subsets of $\calY^{n-m}$ for each given $y^m \in  \calY^m$.  Similarly, for a fixed $y^m$, $\calS_D(\hatV^N(m))\subset\calV^N$ and $\calK_m\subset \calV^N\times\calV^N$ are also deterministic subsets.  Now define the probabilities
\begin{align}
  T_{n,m}(v^N) & :=\bbP\Big(Y_{m+1}^n \in \calQ_n(\hatV^N(m),y^m)\Big| \nn\\*
  &\qquad \qquad V^N=v^N, Y^m=y^m \Big),  \label{eqn:defT}\\
 \overline{T}_{n,m}(v^N)&:=1-T_{n,m}(v^N)  ,\label{eqn:defTbar}\\
L_{n,m}^{(1)}&:=\bbP\Big(Y_{m+1}^n \in \calQ_n(\hatV^N(m),y^m) \Big| \nn\\*
&\qquad\qquad V^N \in \calS_D(\hat{V}^N(m)), Y^m=y^m \Big),\label{eq98most1}\\
L_{n,m}^{(2)}&:=\bbP\Big(Y_{m+1}^n \in \calQ_n(\hatV^N(m),y^m) \Big|\nn\\*
&\qquad\qquad V^N \notin \calS_D(\hat{V}^N(m)), Y^m=y^m \Big),\\
\label{eq98most}
\overline{L}_{n,m}^{(1)}&:=1-L_{n,m}^{(1)},\\
\label{eq99most}
\overline{L}_{n,m}^{(2)}&:=1-L_{n,m}^{(2)}.
\end{align} 
\begin{lemma} \label{lowererror} Fix $m,n \in\bbN$ and $n\geq m$. For the   code fixed in Lemma~\ref{Berlinlem}, the following holds for each $y^m\in\calY^m$:
\begin{align}
\label{pointer2}
&\bbP \left(d(V^N,\hatV^N(n))>D\middle|Y^m=y^m \right)\nonumber\\
&\geq \left(\max\big\{\overline{L}_{n,m}^{(1)},L_{n,m}^{(2)}\big\}\right) \nn\\* 
&\qquad \times \Big(\min\big\{P_{V^N|Y^m}(\calS_D(\hat{V}^N(m))|y^m),\nn\\*  
&\qquad\qquad\qquad\qquad P_{V^N|Y^m}(\overline{\calS}_D(\hat{V}^N(m))|y^m)\big\}\Big ). 
\end{align} 
\end{lemma}
\begin{IEEEproof}
Fix $m\le n$. For each realization $Y^m=y^m$, we have
\begin{align}
&\bbP \left(d(V^N,\hatV^N(n))>D\middle|Y^m=y^m \right)\nn\\
\label{pertin1}
&=\overline{L}_{n,m}^{(1)} P_{V^N|Y^m}(\calS_D(\hat{V}^N(m))|y^m) \nn\\*
&\qquad +  L_{n,m}^{(2)}P_{V^N|Y^m}({\cal\overline{S}}_D(\hat{V}^N(m))|y^m)\\
\label{eq102veryimp}
&\geq \min\big\{P_{V^N|Y^m}(\calS_D(\hat{V}^N(m))|y^m),  \nn\\* 
 &\qquad\qquad  P_{V^N|Y^m}(\overline{\calS}_D(\hat{V}^N(m))|y^m)\big\}  \nn\\* 
&\qquad \times\big(\overline{L}_{n,m}^{(1)}+L_{n,m}^{(2)}\big),
\end{align} \blue{where~\eqref{pertin1} is obtained by writing the excess-distortion probability  given $\{Y^m=y^m\}$ in terms of marginal probabilities.\footnote{Note that~\eqref{pertin1} essentially  follows from the same arguments to obtain~\cite[Unnumbered eq.\ after Eq.~(11)]{Berlin2009a} by setting  $(p_N , p_A)$ in~\cite{Berlin2009a} to be $(P_{V^N|Y^m}(\overline{\calS}_D(\hat{V}^N(m))|y^m), P_{V^N|Y^m}(\calS_D(\hat{V}^N(m))|y^m))$, $\calS $ in~\cite{Berlin2009a}  to be $\calQ_n(\hatV^N(m),y^m)\subset\calY^{n-m}$, and $ (\calQ_N(\calS),\calQ_A(\overline{S}))$ in \cite{Berlin2009a}   to be $(L_{n,m}^{(2)},  \overline{L}_{n,m}^{(1)})$. }
By lower bounding $\overline{L}_{n,m}^{(1)}+L_{n,m}^{(2)}$ in~\eqref{eq102veryimp} by $\max\big\{ \overline{L}_{n,m}^{(1)},L_{n,m}^{(2)}\big\}$, we obtain \eqref{pointer2} and this completes the proof of Lemma~\ref{lowererror}.}
\end{IEEEproof}
\begin{lemma}\label{newsun1lem} 
 Fix $m,n \in\bbN$ and $n\geq m$. For the   code fixed in Lemma~\ref{Berlinlem}, \eqref{newsun1} on the next page holds almost surely:
 \begin{figure*}
\begin{align}
\label{newsun1}
(n-m) B \geq -\bbP(d(\hat{V}^N(n),V^N)\leq D|Y^m) \ln\bigg[ \frac{\bbP(d(\hat{V}^N(n),V^N)>D|Y^m)}{\min\big\{\bbP(d(\hat{V}^N(m),V^N)>D|Y^m),\bbP(d(\hat{V}^N(m),V^N)\leq D|Y^m)\big\}}\bigg]-1.
\end{align}\hrulefill
\end{figure*} 
\end{lemma}
\begin{IEEEproof}
For each fixed $Y^m=y^m$, $\calS_D(\hat{V}^N(m))\subset\calV^N$ is a deterministic set since $\hat{V}^N(m)=g_m(y^m)$. Similarly, $\calK_m\subset\calV^N\times\calV^N$ is also a deterministic set for a  fixed $Y^m=y^m$. It follows that \eqref{eqn:135}--\eqref{eqtrickno1} hold, 
\begin{figure*} 
\begin{align}
L_{n,m}^{(1)}&=\frac{\bbP\big(\{Y_{m+1}^n \in \calQ_n(\hatV^N(m),y^m)\}\cap \{V^N \in \calS_D(\hat{V}^N(m))\} \big|Y^m=y^m \big)}{P_{V^N|Y^m}\big(\calS_D(\hat{V}^N(m))\big|y^m \big)} \label{eqn:135}\\
&=\frac{\sum_{v_1^N \in \calS_D(\hat{V}^N(m))} \bbP\big(\{Y_{m+1}^n \in \calQ_n(\hatV^N(m),y^m)\}\cap \{V^N=v_1^N\}\big|Y^m=y^m \big)}{\sum_{v_1^N\in \calS_D(\hat{V}^N(m))}P_{V^N|Y^m}(v_1^N|y^m)}\\
\label{eqtrickno1}
&=\frac{\sum_{v_1^N \in \calS_D(\hat{V}^N(m))} T_{n,m}(v_1^N) P_{V^N|Y^m}(v_1^N|y^m)}{\sum_{v_1^N\in \calS_D(\hat{V}^N(m))}P_{V^N|Y^m}(v_1^N|y^m)},
\end{align}\hrulefill \end{figure*}
where \eqref{eqtrickno1} follows from Bayes rule and the definition of $T_{n,m}(\cdot)$. 
Similarly, we also have
\begin{align}
\label{eqtrickno2}
L_{n,m}^{(2)}=\frac{\sum_{v_2^N \in \calV^N\setminus \calS_D(\hat{V}^N(m))} T_{n,m}(v_2^N) P_{V^N|Y^m}(v_2^N|y^m)}{\sum_{v_2^N\in \calV^N\setminus\calS_D(\hat{V}^N(m))}P_{V^N|Y^m}(v_2^N|y^m)}.
\end{align}
It follows from~\eqref{eqtrickno1} and~\eqref{eqtrickno2} that
\begin{align}
\label{eqtricky3}
\frac{L_{n,m}^{(1)}}{L_{n,m}^{(2)}} \!=\!\frac{\sum\limits_{(v_1^N,v_2^N) \in \calK_m}\!\!\!\!\!\!\!\!  T_{n,m}(v_1^N) P_{V^N|Y^m}(v_1^N|y^m)P_{V^N|Y^m}(v_2^N|y^m) }{\sum\limits_{(v_1^N,v_2^N)\in\calK_m}\!\!\!\!\!\!\!\!  T_{n,m}(v_2^N) P_{V^N|Y^m}(v_1^N|y^m)P_{V^N|Y^m}(v_2^N|y^m) }.
\end{align} 
Now, from~\eqref{eqtrickno1}, we have \eqref{eqn:140}--\eqref{tricky1}, 
\begin{figure*}
\begin{align}
L_{n,m}^{(1)}\ln \frac{L_{n,m}^{(1)}}{L_{n,m}^{(2)}}&=\left[\frac{\sum_{v_1^N \in \calS_D(\hat{V}^N(m))} T_{n,m}(v_1^N)P_{V^N|Y^m}(v_1^N|y^m)}{P_{V^N|Y^m}({\calS}_D(\hat{V}^N(m))|y^m)}\right]   \left[\frac{\sum_{v_2^N \in \calV^N\setminus \calS_D(\hat{V}^N(m))}P_{V^N|Y^m}(v_2^N|y^m)}{P_{V^N|Y^m}(\overline{\calS}_D(\hat{V}^N(m))|y^m)}\right]\ln\frac{L_{n,m}^{(1)}}{L_{n,m}^{(2)}} \label{eqn:140}\\
&= \frac{ \sum_{(v_1^N,v_2^N) \in \calK_m}P_{V^N|Y^m}(v_1^N|y^m) P_{V^N|Y^m}(v_2^N|y^m)T_{n,m}(v_1^N)\ln\frac{L_{n,m}^{(1)}}{L_{n,m}^{(2)}}}{P_{V^N|Y^m}(\calS_D(\hat{V}^N(m))|y^m)P_{V^N|Y^m}(\overline{\calS}_D(\hat{V}^N(m))|y^m)} \label{eqn:useK}\\
\label{tricky1}
&\leq \frac{\sum_{(v_1^N,v_2^N) \in \calK_m}P_{V^N|Y^m}(v_1^N|y^m)P_{V^N|Y^m}(v_2^N|y^m) T_{n,m}(v_1^N)\ln \frac{T_{n,m}(v_1^N)}{T_{n,m}(v_2^N) }}{P_{V^N|Y^m}(\calS_D(\hat{V}^N(m))|y^m)P_{V^N|Y^m}(\overline{\calS}_D(\hat{V}^N(m))| y^m)},
\end{align}\hrulefill \end{figure*} where~\eqref{eqn:useK} uses the definition $\calK_m$ in~\eqref{eqn:defKm} and~\eqref{tricky1} follows the log-sum inequality and~\eqref{eqtricky3}. 

Similarly, we also have \eqref{tricky2}. 
\begin{figure*}
\begin{align}
\label{tricky2}
\overline{L}_{n,m}^{(1)}\ln \frac{\overline{L}_{n,m}^{(1)}}{\overline{L}_{n,m}^{(2)}}&\leq  \frac{\sum_{(v_1^N,v_2^N) \in \calK_m}P_{V^N|Y^m}(v_1^N|y^m)P_{V^N|Y^m}(v_2^N|y^m) \overline{T}_{n,m}(v_1^N)  \ln \frac{\overline{T}_{n,m}(v_1^N) }{ \overline{T}_{n,m}(v_2^N) }}{P_{V^N|Y^m}(\calS_D(\hat{V}^N(m))|y^m)P_{V^N|Y^m}(\overline{\calS}_D(\hat{V}^N(m))|y^m)}.
\end{align}\hrulefill \end{figure*} 
Now, observe that for every $(v_1^N,v_2^N)\in\calK_m$,
\begin{align}
&T_{n,m}(v_1^N)\ln \frac{T_{n,m}(v_1^N)}{T_{n,m}(v_2^N)}+\overline{T}_{n,m}(v_1^N) \ln \frac{\overline{T}_{n,m}(v_1^N) }{ \overline{T}_{n,m}(v_2^N) }\nonumber\\
\label{eqnew2017}
&\leq \sum_{y_{m+1}^n \in \calY^{n-m}} \bbP\left(Y_{m+1}^n=y_{m+1}^n\big|V^N=v_1^N, Y^m=y^m\right) \nn\\*
&\qquad\times\ln \frac{\bbP\left(Y_{m+1}^n=y_{m+1}^n\big|V^N=v_1^N,Y^m=y^m\right)}{\bbP\left(Y_{m+1}^n=y_{m+1}^n\big|V^N=v_2^N, Y^m=y^m\right)}\\
&=\sum_{y_{m+1}^n \in \calY^{n-m}} \bbP\left(Y_{m+1}^n=y_{m+1}^n\big|V^N=v_1^N, Y^m=y^m\right)\nn\\*
&\qquad\times\sum_{k=m+1}^n \ln \frac{\bbP\left(Y_k=y_k\big|V^N=v_1^N,Y^{k-1}=y^{k-1}\right)}{\bbP\left(Y_k=y_k\big|V^N=v_2^N, Y^{k-1}=y^{k-1}\right)}\label{eqn:memoryless}\\
&=\sum_{k=m+1}^n \!\sum_{y_{m+1}^n \in \calY^{n-m}} \! \! \! \bbP\left(Y_{m+1}^n \! = \! y_{m+1}^n\big|V^N \!= \! v_1^N, Y^m \!= \! y^m\right) \nn\\*
&\qquad\times\ln \frac{\bbP\left(Y_k=y_k\big|V^N=v_1^N,Y^{k-1}=y^{k-1}\right)}{\bbP\left(Y_k=y_k\big|V^N=v_2^N, Y^{k-1}=y^{k-1}\right)}\\
&=\sum_{k=m+1}^n\sum_{y_k \in \calY} \bbP\left(Y_k=y_k\big|V^N=v_1^N, Y^{k-1}=y^{k-1}\right) \nn\\*
&\qquad\times\ln \frac{\bbP\left(Y_k=y_k\big|V^N=v_1^N,Y^{k-1}=y^{k-1}\right)}{\bbP\left(Y_k=y_k\big|V^N=v_2^N, Y^{k-1}=y^{k-1}\right)}\\
\label{keyclosed}
&=\sum_{k=m+1}^n\sum_{y_k \in \calY} P_{Y|X}\left(y_k |f_k(v_1^N, y^{k-1})\right) \nn\\*
&\qquad \times \ln \frac{P_{Y|X}\left(y_k |f_k(v_1^N, y^{k-1}) \right)}{P_{Y|X}\left(y_k |f_k(v_2^N, y^{k-1}) \right)}\\
\label{keypad1}
&\leq (n-m)B.
\end{align}
Here,~\eqref{eqnew2017} follows from the log-sum inequality,~\eqref{eqn:memoryless} follows from the memoryless property of the DMC,~\eqref{keyclosed} follows from $X_k=f_k(V^N,Y^{k-1})$ [cf.~\eqref{eqn:enc}], and~\eqref{keypad1} follows from the definition of $B$ in~\eqref{defineB}.  

Combining~\eqref{tricky1},~\eqref{tricky2}, and~\eqref{keypad1}, we obtain
\begin{align}
\label{eqn:lsi}
L_{n,m}^{(1)} \ln \frac{L_{n,m}^{(1)}}{ L_{n,m}^{(2)}} + \overline{L}_{n,m}^{(1)} \ln \frac{\overline{L}_{n,m}^{(1)}}{\overline{L}_{n,m}^{(2)}}\leq  (n-m)B.
\end{align}
Similarly, we also have 
\begin{align}
\label{eqn:msi}
L_{n,m}^{(2)} \ln \frac{L_{n,m}^{(2)}}{ L_{n,m}^{(1)}} + \overline{L}_{n,m}^{(2)} \ln \frac{\overline{L}_{n,m}^{(2)}}{\overline{L}_{n,m}^{(1)}}\leq  (n-m)B.
\end{align}
Observe that 
\begin{align}
&\bbP(d(\hat{V}^N(n),V^N)\leq D|Y^m=y^m) \nn\\*
\label{newway3}
&=L_{n,m}^{(1)} P_{V^N|Y^m}\big(\calS_D(\hat{V}^N(m))\big|Y^m=y^m\big) \nn\\*
&\qquad+\overline{L}_{n,m}^{(2)} P_{V^N|Y^m}\big(\overline{\calS}_D(\hatV^N(m))\big|Y^m=y^m \big), 
\end{align}
where~\eqref{newway3} follows from~\eqref{pertin1}. 

It follows from~\eqref{eqn:lsi} that
\begin{align}
\label{sim1}
(n-m) B &\geq  -L_{n,m}^{(1)}\ln L_{n,m}^{(2)}-h\big(L_{n,m}^{(1)}\big)\\
\label{term1}
&\geq -L_{n,m}^{(1)} \ln (\max\{\overline{L}_{n,m}^{(1)},L_{n,m}^{(2)}\})-1.
\end{align}
In addition,  it follows  from~\eqref{eqn:msi} that
\begin{align}
(n-m) B &\geq -\overline{L}_{n,m}^{(2)}\ln \overline{L}_{n,m}^{(1)}-h\big(L_{n,m}^{(2)}\big)\\
\label{term2}
&\geq -\overline{L}_{n,m}^{(2)} \ln (\max\{\overline{L}_{n,m}^{(1)},L_{n,m}^{(2)}\})-1.
\end{align}
From~\eqref{term1} and~\eqref{term2} we obtain \eqref{beau_2}--\eqref{beau3}
\begin{figure*}
\begin{align}
(n-m)B  &\geq  -\big[L_{n,m}^{(1)}P_{V^N|Y^m}(\calS_D(\hatV^N(m))|y^m)+ \overline{L}_{n,m}^{(2)}P_{V^N|Y^m}(\overline{\calS}_D(\hatV^N(m))|y^m) \big]\ln (\max\{\overline{L}_{n,m}^{(1)},L_{n,m}^{(2)}\})-1\label{beau_2} \\
\label{beau2}
&=-\bbP(d(\hat{V}^N(n),\hatV^N)\leq D|Y^m=y^m)\ln (\max\{\overline{L}_{n,m}^{(1)},\overline{L}_{n,m}^{(2)}\})-1\\
\label{beau3}
&\geq -\bbP(d(\hat{V}^N(n),\hatV^N)\leq D|Y^m=y^m) \nn\\*
&\qquad\times \ln\bigg[\frac{\bbP(d(\hat{V}^N(n),\hatV^N)> D|Y^m=y^m)}{\min\{\bbP(d(\hat{V}^N(m),V^N)>D|Y^m=y^m),\bbP(d(\hat{V}^N(m),V^N)\leq D|Y^m=y^m)\}}\bigg]-1, 
\end{align}\hrulefill\end{figure*}
where~\eqref{beau2} follows from~\eqref{newway3}, and~\eqref{beau3} follows from Lemma~\ref{lowererror}. This concludes the proof of Lemma~\ref{newsun1lem}.
\end{IEEEproof}

\section{Proof of Lemma \ref{Berlinlem}}\label{proof:Berlinlem}
\begin{IEEEproof}  The proof is partly based on the proof of~\cite[Lemma 1]{Berlin2009a}. For brevity, for $l, m \in \bbN\cup\{0\}$, define
\begin{align}
G_{l,m} &:=\bbP(d(\hat{V}^N(l),V^N)>D|Y^m) \label{eqn:defG}\\
\overline{G}_{l,m} &:= \bbP(d(\hat{V}^N(l),V^N)\leq D|Y^m) = 1- G_{l,m} \label{eqn:defF}\\
\Lambda_m &:=-\ln\left[ \min\left\{G_{m,m},\overline{G}_{m,m}\right\}\right].
\end{align}
Note that if $m=0$ in~\eqref{eqn:defG} or~\eqref{eqn:defF}, we drop the conditioning on $Y^m$ in the probabilities and thus $G_{l,0}$ and $\overline{G}_{l,0}$ are deterministic.
From the definitions of $\tau_N'$ and $\tau_N^*$ in~\eqref{definetau'} and~\eqref{definetau^*} respectively, we have $\tau_N'\geq \tau_N^*$. Hence from Lemma~\ref{newsun1lem} (with $n$ replaced by $\tau_N'\wedge n$ and $m$ replaced by $\tau_N^*\wedge n$), we have  for all $n\in \bbN$,
\begin{align}
&(\tau_N'\wedge n-\tau_N^*\wedge n) B \nonumber\\
&\geq -\overline{G}_{\tau_N'\wedge n ,\tau_N^*\wedge n }\ln\bigg[ \frac{G_{\tau_N'\wedge n ,\tau_N^*\wedge n }}{\min \{G_{\tau_N*\wedge n ,\tau_N^*\wedge n },\overline{G}_{\tau_N*\wedge n ,\tau_N^*\wedge n }\}}\bigg]  \nn\\*
&\qquad\qquad-1, \quad \mbox{a.s.}\label{eqtest}
\end{align}

On the other hand, from~\eqref{finding} we have 
\begin{align}
\label{keyresp}
\bbP(\tau_N^*<\infty)=\bbP(\tau_N'<\infty)=1,
\end{align} 
hence, by the definitions of $\tau_N^*$ in~\eqref{definetau^*}, $\tau_N'$ in~\eqref{definetau'}, and the fact in~\eqref{keyresp}, the following inequalities hold almost surely:
\begin{align}
\label{eq1app}
\!\!\min_{v^N \in \calV^N}\!\!\sum_{w^N \in \calV^N\setminus \calS_D(v^N)} P_{V^N|Y^{\tau_N^*}}(w^N|Y^{\tau_N^*})&\leq \delta_N,\\
\label{eq2app}
\!\!\min_{v^N \in \calV^N}\!\! \sum_{w^N \in \calV^N\setminus \calS_D(v^N)} P_{V^N|Y^{\tau_N^*-1}}(w^N|Y^{\tau_N^*-1})&>\delta_N.
\end{align}
It follows from~\eqref{eq2app} and Lemma~\ref{notepoint} that
\begin{align}
\label{eq3app}
\!\min_{v^N \in \calV^N} \sum_{w^N \in \calV^N\setminus \calS_D(v^N)} \!\! P_{V^N|Y^{\tau_N^*}}(w^N|Y^{\tau_N^*})\! >\!\lambda \delta_N, \;\mbox{a.s.}
\end{align}
Since we use the distortion-MAP decoding at time $\tau_N^*$, by~\eqref{keyfact},~\eqref{eq1app}, and~\eqref{eq3app}, 
\begin{align}
\delta_N \geq G_{\tau_N^* ,\tau_N^*} > \lambda \delta_N, \quad \mbox{a.s.}
\end{align}
Hence, we have 
\begin{align}
\label{eq4app}
\min\left\{G_{\tau_N^* ,\tau_N^*} ,\overline{G}_{\tau_N^* ,\tau_N^*} \right\} \geq \min\{\lambda \delta_N, 1-\delta_N\}, \quad \mbox{a.s.}
\end{align}
Now, observe that
\begin{align}
\label{dominate1}
&\lim_{n\to \infty}G_{\tau_N^*\wedge n ,\tau_N^*\wedge n  }\nn\\*
&=\lim_{n\to \infty}\min_{v^N \in \calV^N}\sum_{w^N \in \calV^N\setminus \calS_D(v^N)} P_{V^N|Y^{\tau_N^*\wedge n}}(w^N|Y^{\tau_N^*\wedge n})\\
\label{contmap1}
&= \min_{v^N \in \calV^N} \lim_{n\to \infty} \sum_{w^N \in \calV^N\setminus \calS_D(v^N)} P_{V^N|Y^{\tau_N^*\wedge n}}(w^N|Y^{\tau_N^*\wedge n}) \\
&=\min_{v^N \in \calV^N}\sum_{w^N \in \calV^N\setminus \calS_D(v^N)} \lim_{n\to \infty} P_{V^N|Y^{\tau_N^*\wedge n}}(w^N|Y^{\tau_N^*\wedge n}) \\
\label{contmap2}
&=\min_{v^N \in \calV^N}\sum_{w^N \in \calV^N\setminus \calS_D(v^N)}P_{V^N|Y^{\tau_N^*}}(w^N|Y^{\tau_N^*}),\\
&= G_{\tau_N^*,\tau_N^*}. \label{eq135h}
\end{align}
Here,~\eqref{contmap1} follows from $\lim_{n\to \infty} \min\{X_n,Y_n\}=\min\{X,Y\}$ a.s. if $\lim_{n\to \infty} X_n=X$ a.s. and $\lim_{n\to \infty} Y_n=Y$ a.s.,~\eqref{contmap2} follows from   L\'{e}v{y}'s zero-one law~\cite{Durrett} since $\{\sigma(Y^{\tau_N^*\wedge n})\}_{n=1}^{\infty}$ is a filtration and $\sigma(Y^{\tau_N^*})$ is the maximal $\sigma$-algebra generated by  $\{\sigma(Y^{\tau_N^*\wedge n})\}_{n=1}^{\infty}$. It follows from~\eqref{eqn:defF} and~\eqref{eq135h} that
\begin{equation}
\lim_{n\to\infty}\overline{G}_{\tau_N^*\wedge n,\tau_N^*\wedge n}=\overline{G}_{\tau_N^*,\tau_N^*}.\label{eq138h}
\end{equation}
Hence by taking $n$ to infinity in~\eqref{eqtest} and applying~\eqref{eq135h} and~\eqref{eq138h}, we have 
\begin{align}
\Gamma_N&:=\liminf_{n\to \infty}\Big[\overline{G}_{\tau_N'\wedge n,\tau_N^*\wedge n}\Lambda_{\tau_N^*\wedge n }\nonumber\\
&\qquad+(\tau_N'\wedge n-\tau_N^*\wedge n) B+ 1\nn\\*
&\qquad+\overline{G}_{\tau_N'\wedge n,\tau_N^*\wedge n} \ln G_{\tau_N^*\wedge n,\tau_N^*\wedge n} \Big] \label{eqn:defGamma_N}\\
\label{eq142h}
&\leq \limsup_{n\to \infty} \Big[\overline{G}_{\tau_N'\wedge n,\tau_N^*\wedge n}\Lambda_{\tau_N^*\wedge n }\Big]\nonumber\\
&\qquad+\liminf_{n\to \infty}\Big[ (\tau_N'\wedge n-\tau_N^*\wedge n) B+ 1 \nn\\*
&\qquad+\overline{G}_{\tau_N'\wedge n,\tau_N^*\wedge n} \ln G_{\tau_N'\wedge n,\tau_N^*\wedge n} \Big]  \\
&\leq -\ln\left[ \min\{\lambda \delta_N, 1-\delta_N\}\right]\nonumber\\
&\qquad+\liminf_{n\to \infty}\Big[ (\tau_N'\wedge n-\tau_N^*\wedge n) B+ 1\nn\\*
&\qquad+\overline{G}_{\tau_N'\wedge n,\tau_N^*\wedge n} \ln G_{\tau_N'\wedge n,\tau_N^*\wedge n} \Big]\label{eq142h2}, 
\end{align}
where for~\eqref{eq142h2}, we note that  the first term in \eqref{eq142h} can be upper bounded by $-\ln\left[ \min\{\lambda \delta_N, 1-\delta_N\}\right]$ because  $\overline{G}_{\tau_N'\wedge n,\tau_N^*\wedge n}\le 1$ and $\Lambda_{\tau_N^*\wedge n }\ge 0$  as well as~\eqref{eq135h},~\eqref{eq138h}, and~\eqref{eq4app}. 

Now we record a simple fact that follows from the bounded convergence theorem. We have 
\begin{align}
&\lim_{n\to \infty}G_{\tau_N'\wedge n,0}  \nn\\*
& =\lim_{n\to \infty}\bbP(d(\hat{V}^N(\tau_N'\wedge n),V^N)>D)\\
&=\lim_{n\to \infty}\bbE \big[\bone\{d(\hat{V}^N(\tau_N'\wedge n), V^N)> D\} \big]\\
&=\bbE\Big[\lim_{n\to \infty}\bone\{d(\hat{V}^N(\tau_N'\wedge n), V^N)> D\} \Big]\\
&=\bbE\big[\bone\{d(\hat{V}^N(\tau_N'), V^N)> D)\}\big]\\
&=\bbP(d(\hat{V}^N(\tau_N'),V^N)>D)= G_{\tau_N',0} . \label{eqn:bct}
\end{align}
It follows by taking expectations on both sides  of \eqref{eq142h2}  that
\begin{align}
&\bbE[\Gamma_N] \nn\\*
&\leq -\ln\left[ \min\{\lambda \delta_N, 1-\delta_N\}\right] \nonumber\\
&\qquad+\bbE\Big[\liminf_{n\to \infty}\big[ (\tau_N'\wedge n-\tau_N^*\wedge n) B+ 1\nn\\*
&\qquad+\overline{G}_{\tau_N'\wedge n,\tau_N^*\wedge n} \ln G_{\tau_N'\wedge n,\tau_N^*\wedge n} \big]\Big]\\
\label{eq149hn}
&= -\ln\left[ \min\{\lambda \delta_N, 1-\delta_N\}\right] \nonumber\\
&\qquad+\bbE\Big[\lim_{n\to \infty}\big[ (\tau_N'\wedge n-\tau_N^*\wedge n) B\big]\Big]\nn\\*
&\qquad+ \bbE\Big[\liminf_{n\to \infty}\big[ 1 +\overline{G}_{\tau_N'\wedge n,\tau_N^*\wedge n} \ln G_{\tau_N'\wedge n,\tau_N^*\wedge n}\big]\Big]\\
\label{eq150hn}
&\leq -\ln\left[ \min\{\lambda \delta_N, 1-\delta_N\}\right] \nonumber\\
&\qquad+\bbE\Big[\lim_{n\to \infty}\big[ (\tau_N'\wedge n-\tau_N^*\wedge n) B\big]\Big] \nn\\*
&\qquad+\liminf_{n\to \infty}\bbE  \Big[ 1 +\overline{G}_{\tau_N'\wedge n,\tau_N^*\wedge n} \ln \big( 1-\overline{G}_{\tau_N'\wedge n,\tau_N^*\wedge n} \big)\Big] \\
\label{eq151hn}
&\leq -\ln\left[\min\{\lambda \delta_N, 1-\delta_N\}\right] +\bbE[\tau_N'-\tau_N^*] B\nn\\*
&\qquad+\liminf_{n\to \infty}\Big[1+ \overline{G}_{\tau_N'\wedge n,0} \ln \big( 1-\overline{G}_{\tau_N'\wedge n,0}\big)\Big]\\
\label{eq152hn}
&= -\ln\left[ \min\{\lambda \delta_N, 1-\delta_N\}\right]  +\bbE[\tau_N'-\tau_N^*] B+1\nn\\*
&\qquad+ \overline{G}_{\tau_N' ,0} \ln \big( 1-\overline{G}_{\tau_N',0}\big)\\
&= -\ln\left[ \min\{\lambda \delta_N, 1-\delta_N\}\right]  +\bbE[\tau_N'-\tau_N^*] B+1\nn\\*
&\qquad+ (1-{G}_{\tau_N' ,0}) \ln \big( {G}_{\tau_N',0}\big)\\
\label{supsample3}
&\leq - \ln\left[ \min\{\lambda \delta_N, 1-\delta_N\}\right]+ \bbE[\tau_N'-\tau_N^*] B+  1 \nn\\*
&\qquad+ (1-\rvP_{\rmd}(N,D))\ln \rvP_{\rmd}(N,D).
\end{align}
Here,
~\eqref{eq150hn} follows from Fatou's lemma,~\eqref{eq151hn} follows from the fact that $x\in(0,1)\mapsto x \ln (1-x)$ is concave and the law of iterated expectations so $\bbE[\overline{G}_{\tau_N'\wedge n,\tau_N^*\wedge n} ] = \overline{G}_{\tau_N'\wedge n,0}$ and $\bbE[G_{\tau_N'\wedge n,\tau_N^*\wedge n} ] =G_{\tau_N'\wedge n,0}$,~\eqref{eq152hn} follows from~\eqref{eqn:bct},
and finally~\eqref{supsample3} follows from~\eqref{eqcompare} of Lemma~\ref{stopcomp} and  the increasing nature of the function $(1-x) \ln x$ in $0<x<1$.

Hence, by the definition of $\Gamma_N$ in \eqref{eqn:defGamma_N}, the fact that $\Gamma_N\ge 0$ a.s.\ in \eqref{eqtest}, and the bound in~\eqref{supsample3}, we have that 
\begin{align}
0  \leq \bbE[\Gamma_N] 
\label{supsample100}
&\leq \bbE[\tau_N'-\tau_N^*] B - \ln\left[ \min\{\lambda \delta_N, 1-\delta_N\}\right] \nn\\*
&\qquad+1+(1-\rvP_{\rmd}(N,D))\ln \rvP_{\rmd}(N,D).
\end{align}
Because $-\rvP_{\rmd}(N,D)\ln \rvP_{\rmd}(N,D)\le 1$, 
 it follows that
\begin{align}
\bbE[\tau_N'-\tau_N^*] B + \ln \rvP_{\rmd}(N,D)+2 \geq \ln\left[ \min\left\{\lambda\delta_N,1-\delta_N\right\}\right],
\end{align}
or equivalently,~\eqref{eq84best}. This concludes the proof of Lemma~\ref{Berlinlem}.
\end{IEEEproof}
\subsection*{Acknowledgements} 
The authors would   like to thank the Associate Editor Prof.\ Mich\`ele Wigger and the anonymous reviewers for exceedingly helpful and extensive feedback that helped to   improve the presentation in the paper.

The authors would  also like to   thank  Professors Hideki Yagi, Shun Watanabe, and Tetsunao Matsuta for suggestions concerning  the  optimality of the MAP-decoding rule, leading to the proof of Lemma~\ref{key}. Finally, the authors would like to   thank Prof.\ Baris Nakibo\u{g}lu for helpful discussions during the initial phase of this work. 

The authors are supported in part by an NUS-JSPS grant under grant number R-263-000-C57-133/114.


\bibliographystyle{unsrt}
\bibliography{isitbib}

\begin{IEEEbiographynophoto}{Lan V. Truong} (S'12--M'15) received the B.S.E.\ degree in electronics and telecommunications from the Posts and Telecommunications Institute of Technology (PTIT), Hanoi, Vietnam, in 2003, and the M.S.E.\ degree from the School of Electrical and Computer Engineering, Purdue University, West Lafayette, IN, USA, in 2011, and the Ph.D.\ degree from the Department of Electrical and Computer Engineering, National University of Singapore (NUS), Singapore, in 2018. He was an Operation and Maintenance Engineer with MobiFone Telecommunications Corporation, Hanoi, for several years. He spent one year as a Research Assistant with the NSF Center for Science of Information and Department of Computer Science, Purdue University, in 2012. From 2013 to 2015, he was an Academic Lecturer with the Department of Information Technology Specialization, FPT University, Hanoi, Vietnam. Since 2018, he has been a Research Assistant/Post-Doctoral Research Fellow with the Department of Computer Science, School of Computing, NUS. His research interests include information theory, machine learning, and communications.  
\end{IEEEbiographynophoto}

\begin{IEEEbiographynophoto}{Vincent Y.\ F.\ Tan} (S'07-M'11-SM'15)  was born in Singapore in 1981. He is currently a Dean's Chair Associate Professor in the Department of Electrical and Computer Engineering  and the Department of Mathematics at the National University of Singapore (NUS). He received the B.A.\ and M.Eng.\ degrees in Electrical and Information Sciences from Cambridge University in 2005 and the Ph.D.\ degree in Electrical Engineering and Computer Science (EECS) from the Massachusetts Institute of Technology (MIT)  in 2011.  His research interests include information theory, machine learning, and statistical signal processing.

Dr.\ Tan received the MIT EECS Jin-Au Kong outstanding doctoral thesis prize in 2011, the NUS Young Investigator Award in 2014,  the Singapore National Research Foundation (NRF) Fellowship (Class of 2018) and the NUS Young Researcher Award in 2019. He is also an IEEE Information Theory Society Distinguished Lecturer for 2018/9. He has authored a research monograph on {\em ``Asymptotic Estimates in Information Theory with Non-Vanishing Error Probabilities''} in the Foundations and Trends in Communications and Information Theory Series (NOW Publishers). He is currently serving as an Associate Editor of the IEEE Transactions on Signal Processing.
\end{IEEEbiographynophoto}

\end{document}